\documentclass[]{pasj01}
\usepackage{times}

\Received{2020/12/23}
\Accepted{2021/06/11}
\Published{yyyy/mm/dd}

\newcommand{\bey}{\begin{eqnarray}}
\newcommand{\eey}{\end{eqnarray}}
\newcommand{\be}{\begin{equation}}
\newcommand{\ee}{\end{equation}}

\newcommand{\nn}{\nonumber}

\newcommand{\ihmpct}{\, h^3\, {\rm Mpc}^{-3}}

\newcommand{\wt}{\widetilde}

\begin{document}

\title{Angular clustering and host halo properties of [OII] emitters at $z >1$ in the Subaru HSC survey 
}

\author{
Teppei \textsc{Okumura},\altaffilmark{1,2}${}^*$  
Masao \textsc{Hayashi},\altaffilmark{3} 
I-Non \textsc{Chiu},\altaffilmark{1} 
Yen-Ting \textsc{Lin},\altaffilmark{1}  \\
Ken \textsc{Osato},\altaffilmark{4}
Bau-Ching \textsc{Hsieh},\altaffilmark{1} 
and
Sheng-Chieh \textsc{Lin}\altaffilmark{1,5} 
}

\email{tokumura@asiaa.sinica.edu.tw}

\altaffiltext{1} {Institute of Astronomy and Astrophysics, Academia Sinica, No. 1, Section 4, Roosevelt Road, Taipei 10617, Taiwan}
\altaffiltext{2}{Kavli Institute for the Physics and Mathematics of the Universe (WPI), UTIAS, The University of Tokyo, Kashiwa, Chiba 277-8583, Japan}
\altaffiltext{3} {National Astronomical Observatory, Mitaka, Tokyo 181-8588, Japan}
\altaffiltext{4}{Institut d'Astrophysique de Paris, Sorbonne Universit\'e, CNRS, UMR 7095, 75014 Paris, France}
\altaffiltext{5}{Department of Physics and Astronomy, University of Kentucky, 505 Rose Street, Lexington, KY 40506, USA.}


\maketitle

\begin{abstract}
We study the angular correlation function of star-forming galaxies and
properties of their host dark matter halos at $z>1$ using the
Hyper-Suprime Cam (HSC) Subaru Strategic Program (SSP) survey.  We use [OII] emitters identified
using two narrow-band (NB) filters, {\it NB816} and {\it NB921}, in
the Deep/UltraDeep layers, which respectively cover large angular
areas of $16.3 ~{\rm deg}^2$ and $16.9~{\rm deg}^2$.  Our sample
contains 8302 and 9578 [OII] emitters at $z=1.19$ ({\it NB816}) and
$z=1.47$ ({\it NB921}), respectively.  We detect a strong clustering
signal over a wide angular range, $0.001 < \theta < 1$ [deg], with the
bias $b=1.61^{+0.13}_{-0.11}$ ($z=1.19$) and $b=2.09^{+0.17}_{-0.15}$
($z=1.47$). We also find a clear deviation of the correlation from a
simple power-law form.  To interpret the measured clustering signal,
we adopt a halo occupation distribution (HOD) model that is
constructed to explain the spatial distribution of galaxies selected
by a star formation rate.  The observed correlation function and
number density are simultaneously explained by the best-fitting HOD
model.  From the constrained HOD model, the average mass of halos
hosting the [OII] emitters is derived to be $\log{M_{\rm
    eff}/(h^{-1}M_\odot)}=12.70^{+0.09}_{-0.07}$ and
$12.61^{+0.09}_{-0.05}$ at $z=1.19$ and $1.47$, respectively, which
will become halos with the present-day mass, $M\sim 1.5 \times
10^{13}h^{-1}M_\odot$.  The satellite fraction of the [OII] emitter
sample is found to be $f_{\rm sat}\sim 0.15$.  All these values are
consistent with the previous studies of similar samples, but we obtain
tighter constraints  
even in a larger parameter space due to the larger sample size from the HSC.
The results obtained for host halos of [OII] emitters in this paper 
enable the construction of mock galaxy catalogs  
and the systematic forecast study of cosmological constraints 
from upcoming emission line galaxy surveys 
such as the Subaru Prime Focus Spectrograph survey.
\end{abstract}

\section{Introduction}\label{sec:introduction}

Observation of the large-scale structure of the Universe via the
distribution of galaxies in galaxy redshift surveys provides a
powerful tool to reveal the nature of dark matter and dark energy
\citep{Peebles:1980,Weinberg:2013}.  As dark energy started to become
a dominant energy component toward today, the Universe experienced a
transition from decelerating to accelerating expansion at redshift
around unity.  It is thus crucial to investigate the large-scale
spatial distribution of galaxies over a broad redshift range from
$z=0$ to $z>1$.

The relation between galaxies and their host dark matter halos has
been extensively investigated using the prescription of the halo
occupation distribution (HOD) modeling
\citep{Jing:1998a,Seljak:2000,Peacock:2000,Ma:2000,Scoccimarro:2001,Berlind:2002,Cooray:2002}.
Galaxies selected by the colors, specific star formation rates (sSFR), and stellar masses, have been mainly used as tracers of the
large-scale structure in wide-angle galaxy redshift surveys.  There
are a lot of preceding studies of the HOD modeling for such
populations, and it is known that a simple model proposed by
\citet{Zheng:2005} can describe the spatial distribution of these
galaxies (e.g.,
\cite{Zehavi:2005,Blake:2008,Zheng:2009,Abbas:2010,Zehavi:2011,White:2011,Wake:2011,Coupon:2012,de_la_Torre:2013,Reid:2014,Guo:2014,Koda:2016,Ishikawa:2020}).
Most of the work is, however, limited at redshift less than unity
because these types of galaxies 
become harder to target at higher redshifts \citep{Takada:2014,DESI-Collaboration:2016}:
for early/red-type galaxies one needs long exposure time to get the $4000$\AA ~  break and 
for color-selected galaxies the accuracy of the photometric redshift becomes worse at such redshifts. 
On the other hand, statistical properties of dark matter
halos have been studied using the HOD of high-$z$ quasars and
galaxies, i.e., Lyman-break galaxies or Lyman-$\alpha$ emitters at
$z>2$, for relatively narrow angular regions
\citep{Bullock:2002,Moustakas:2002,Hamana:2004,Conroy:2006,Richardson:2012,Kayo:2012,Durkalec:2015,Harikane:2016,Ishikawa:2017,Harikane:2018}.

Due to the observational limit of color/sSFR/stellar mass-selected galaxies at higher redshifts, recently the focus of large galaxy surveys has
turned to emission line galaxies (ELGs) that can be targeted 
at $z>1$. Such surveys include the Subaru
FastSound survey \citep{Tonegawa:2015,Okada:2016}, extended Baryon
Oscillation Spectroscopic Survey (eBOSS: \cite{Dawson:2016}), the
Hobby-Eberly Telescope Dark Energy Experiment (HETDEX:
\cite{Adams:2011}), Subaru Prime Focus Spectrograph (PFS:
\cite{Takada:2014}), Dark Energy Spectroscopic Instrument (DESI:
\cite{DESI-Collaboration:2016}), the Spectro-Photometer for the
History of the Universe, Epoch of Reionization, and Ices Explorer
(SPHEREx: \cite{Dore:2014}), and {\it Euclid} \citep{Laureijs:2011}.
However, statistical properties of the host halos for ELGs are not as simple as those
for galaxies selected by colors, sSFR and stellar masses because unlike them 
there does not necessarily exist an ELG in the center of a halo at the massive end.  \citet{Geach:2012} proposed a
model of HOD for ELGs taking into account the fact that ELGs selected
via star formation rates do not necessarily reside in the centers of
massive halos. Then they placed a constraint on the HOD parameters for
H$\alpha$ emitters from the Hi-Z Emission Line Survey (HiZELS:
\cite{Geach:2008}).  The HOD of the HiZELS sample has been reanalyzed
by \citet{Cochrane:2017} and \citet{Cochrane:2018}.  Although a simple
HOD modeling has been performed for H$\alpha$ emitters of the
FastSound survey at $z\sim 1.4$ in \citet{Okumura:2016}, the sample
was so sparse that the clustering signal has been consistent with the
HOD of central galaxies only.  \citet{Kashino:2017a} analyzed the
H$\alpha$-selected galaxies from the FMOS-COSMOS survey at $z\sim 1.6$
and constrained the HOD model, but they still adopted the simple model
of \citet{Zheng:2005}.  The detailed HOD modeling has been performed
by \citet{Hong:2019} for Ly$\alpha$ emitters at $z\sim 2.67$ selected
from the NOAO Deep Wide-Field Survey.

Since [OII] emitters ($\lambda =$ 3726, 3729 \AA) are one of the main tracers of the large-scale
structure employed in upcoming cosmological surveys, it is of crucial
importance to investigate properties of halos which host them.  So
far, however, there are few observational studies of the HOD modeling
for [OII] emitters.  From the HiZELS survey \citet{Khostovan:2018}
measured the clustering of H$\beta$ + [OIII] as well as [OII]
emitters, but they did not find deviation of the correlation function
from a power-law form.  Rather than the HOD, properties of [OII]
emitters have been investigated using a semi-analytic model of galaxy
formation (e.g.,
\cite{Contreras:2013b,Gonzalez-Perez:2018,Favole:2020,Gonzalez-Perez:2020,Avila:2020}).
\citet{Guo:2019} analyzed the clustering of
spectroscopically-identified [OII] emitting galaxies at $0.7 < z <
1.2$ from the eBOSS survey.  They then investigated properties of the
host halos using the conditional stellar mass function.  Recently
there are also attempts to determine the HOD of ELGs by directly
identifying emission lines in cosmological hydrodynamical simulations
\citep{Hadzhiyska:2020,Osato:2021}.  Properties of the luminosity
function of [OII] and other ELGs are also being actively investigated (e.g.,
\cite{Comparat:2016,Saito:2020,Hayashi:2020,Gao:2020}).

\begin{table*}[bt!]
\caption{Summary of the [OII] ELG data used in this paper.${}^*$}
\begin{center}
\begin{tabular}{l | cccc | cccc}
\hline 
&  \multicolumn{4}{c|}{{\it NB816} ($1.178<z<1.208$)}   &  \multicolumn{4}{c}{{\it NB921} ($1.453<z<1.489$)}   \\ 
Field &  Area (${\rm deg}^2$) & $N_g $ & $n_g (\ihmpct)$ & $N_{\rm sub}$ & Area (${\rm deg}^2$) & $N_g$ & $n_g(\ihmpct)$& $N_{\rm sub}$ \\
\hline
COSMOS & 1.37 & 566 &$4.48 \times 10^{-3}$ &9 & 5.80 & 3134 & $4.30 \times 10^{-3}$  & 41 \\        
DEEP2-3 & $4.98$ & 2485 & $5.40 \times 10^{-3}$ & 35 & 4.92 & 2107 & $3.41 \times 10^{-3}$ & 35\\   
ELAIS-N1 & $4.81$ & 2927 & $6.59 \times 10^{-3}$ & 34 & 4.81 & 3098 & $5.13 \times 10^{-3}$ & 34\\ 
SXDS+XMM-LSS & $5.17$  & 2324 & $4.86 \times 10^{-3}$ & 37 & 1.33 & 1239 & $7.44 \times 10^{-3}$ & 9\\
\hline
Total & $16.33$ & 8302 & $5.50 \times 10^{-3}$& 115 & 16.86 & 9578 & $4.53 \times 10^{-3}$ & 119\\ 
\hline
\end{tabular}
\end{center}
\label{tab:four_fields}
\begin{tabnote}
$*$ The value of the area shows the one after masking with bright
  object masks, $N_g$ is the number of the galaxies used in this paper
  after making the magnitude and flux cuts, $n_g$ is the number
  densities of the galaxies, and $N_{\rm sub}$ is the number of
  sub-regions using jackknife resampling (see section
  \ref{sec:covariance}). The last row shows the values for the sum of
  the four fields.
\end{tabnote}
\end{table*}

In this paper we present a detailed HOD modeling for [OII] emitters
identified by Narrow-Band (NB) filters in the Subaru Hyper-Suprime Cam
(HSC) survey.  NB imaging surveys of ELGs allow us to cover a wide and
homogeneous field of view.  Thus it provides a suitable sample to
probe the large-scale spatial distribution of emission-line galaxies.
We, for the first time, constrain the general HOD model proposed by
\citet{Geach:2012} for populations of [OII] emitters at $z=1.19$ and
$z=1.47$ in the HSC survey.  We then discuss properties of dark matter
halos which host the [OII] emitters based on the constrained HOD
parameters.

The structure of this paper is as follows.  Section \ref{sec:data}
describes the HSC survey and the [OII] emitter sample.  Section
\ref{sec:measurement} presents the measurements of the angular
correlation function and its covariance matrix.  Power-law and
linearly-biased dark matter model fittings are shown in section
\ref{sec:result}. We also discuss dependences of  
the correlation function amplitude on different stellar mass and emission line luminosity thresholds.
The detailed HOD modeling is performed for the
measured correlation functions in section \ref{sec:hod}.  Conclusions
are given in section \ref{sec:conclusion}.  We compare the angular
correlation functions measured from four individual fields in appendix
\ref{sec:w_individual}. 
We present the HOD parameter constraints without using the information of 
the measured abundance of [OII] emitters in appendix \ref{sec:hod_wo_nobs}. 

Throughout this paper, we assume a flat $\Lambda$CDM cosmology with
the cosmological parameters based on the results of Planck CMB
measurements \citep{Planck-Collaboration:2016}:
$\Omega_m=1-\Omega_\Lambda = 0.307$, $\Omega_b=0.0486$, $h=0.677$,
$n_s=0.967$ and $\sigma_8=0.816$.

\section{Data} \label{sec:data}

\subsection{Hyper-Suprime Cam Survey} \label{sec:hsc}

The HSC Subaru Strategic Program (SSP) is an ongoing imaging survey since 2014 \citep{Aihara:2018}
with a $1.77~{\rm deg}^2$ field-of-view imaging camera installed on the Subaru Telescope
\citep{Miyazaki:2012,Miyazaki:2018,Furusawa:2018,Komiyama:2018}.

In the HSC survey the deep imaging is conducted in five broad-band
(BB) filters as well as four Narrow-Band (NB) filters in the Deep (D)
and UltraDeep (UD) layers over $28~{\rm deg}^2$ in total.  The imaging
reduction and catalog construction are carried out by \citet{Bosch:2018}
and \citet{Huang:2018a}.  In 2019, the second public data release
(PDR2) of the HSC SSP data has been made
\citep{Aihara:2019}.

\begin{figure*}[bt]
\begin{center}
\vspace{-.2cm}
\includegraphics[width=1.\textwidth]{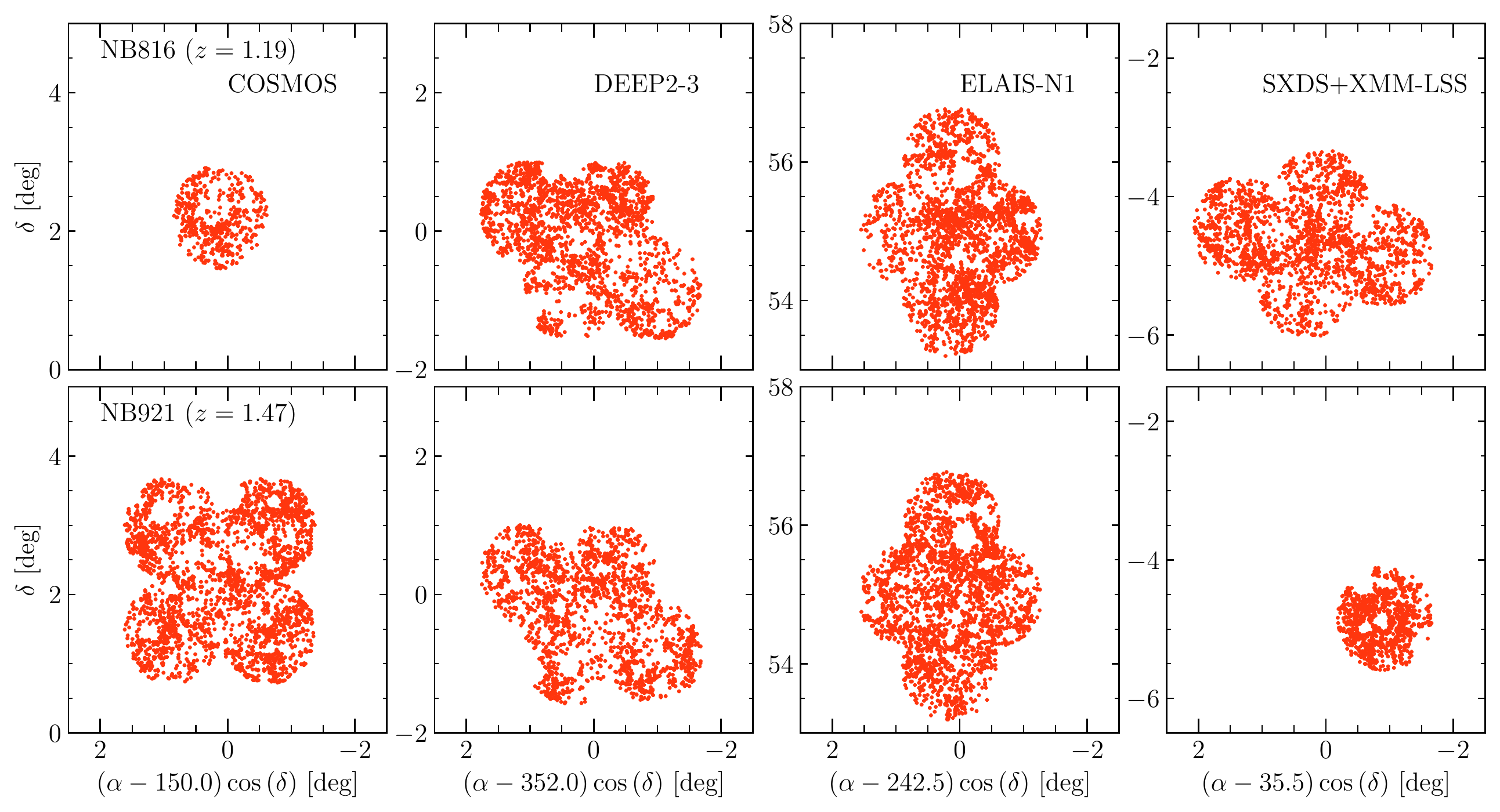}
\end{center}
\caption{Angular distribution of [OII] emitters at $z=1.19$ (top) and $z=1.47$ (bottom) with line flux greater than $3\times
  10^{-17}{\rm erg ~s}^{-1}{\rm cm}^{-2}$ and magnitude
  brighter than $23.5~ {\rm mag}$ in the NB filters, {\it NB816}  and {\it NB921}, 
  respectively, for the individual four fields.  The large empty
  areas where no galaxy is found are mostly the regions masked for
  bright objects.  See figure 4 of \citet{Hayashi:2020} for details
  of the mask information.  }
\label{fig:angular_dist}
\end{figure*}

\subsection{Narrow Band filters and [OII] emitter sample}\label{sec:data_nb}
We use the data from two NB filters, {\it NB816} and {\it NB921},
constructed by \citet{Hayashi:2020} based on the D and UD layers of
Subaru-SSP from the PDR2.  These data are significantly updated from
the version of PDR1 \citep{Aihara:2018a,Hayashi:2018}.  The coverage
of {\it NB816} and {\it NB921} increases to $26~{\rm deg}^2$ covered
by 13 pointings and $28~{\rm deg}^2$ covered by 14 pointings,
respectively, from the PDR1 \citep{Aihara:2019}.  We apply the bright
star masks modified from \citet{Coupon:2018} and \citet{Aihara:2019}
because \citet{Hayashi:2020} noticed that the mask size for some
bright stars is not large enough to remove false detections around 
prominent stellar halos.  As a result, the total area is reduced to
$16.3~{\rm deg}^2$ ({\it NB816}) and $16.9~{\rm deg}^2$ ({\it NB921}),
as shown in table \ref{tab:four_fields}.

The survey area in the D layer, where the NB data are available,
consists of four separate fields: E-COSMOS, DEEP2-3, ELAIS-N1, and
XMM-LSS.  Furthermore, each of E-COSMOS and XMM-LSS fields encompasses the
UD layer covered by a single pointing of HSC, named UD-COSMOS and SXDS,
respectively.  Since UD-COSMOS and E-COSMOS are jointly processed, 
we call the combination the COSMOS field.

The complete description of the [OII] emitter sample used in this study 
is provided in \citet{Hayashi:2018} and \citet{Hayashi:2020}. 
While we briefly summarize our sample in the following, we refer the reader 
to these papers for more details.
The ELGs are selected using the NB data together with data from the five
BB filters. 
[OII] emitters are
observed by {\it NB816} and {\it NB921} filters in two narrow redshift
ranges, $1.178<z<1.208$ and $1.453<z<1.489$, respectively.  
To identify the [OII] emitters, we use spectroscopic redshifts if available and photometric redshifts otherwise.
If a galaxy has a spectroscopic redshift outside of the ranges above, we remove 
it as a contaminant. When the photometric redshift is used, we have taken into account 
its uncertainty for the identification of [OII] emitters. 
For galaxies not identified with spectroscopic or photometric redshifts, 
we use the color-color diagrams to distinguish [OII] emitters from 
other possibilities. 
Galaxies selected as NB [OII] emitters are expected to be star-forming galaxies \citep{Ly:2012,Hayashi:2013}. 
Thus, our sample is limited by star forming rate (SFR). 
While our [OII] emission lines are affected by the dust extinction, 
the majority of typical star-forming galaxies is properly selected
because the cosmic SFR density in our catalogs is consistent with other preceding studies
\citep{Hayashi:2020}.

We make magnitude and flux cuts 
on our data, the model magnitude\footnote{We use a composite model magnitude
named \texttt{cmodel}.} $m\leq 23.5$ and the estimated line flux of $\geq
3\times 10^{-17}{\rm erg ~s}^{-1}{\rm cm}^{-2}$. We refer the reader to \citet{Hayashi:2020} for the details of the line flux estimation.
The limiting magnitude of the NB filters is much deeper than $23.5$ [mag] so that fainter [OII] emitters have been included in our sample, and accordingly [OII] emitters which have the line flux with $(1-2)\times 10^{-17}{\rm erg ~s}^{-1}{\rm cm}^{-2}$ have been selected. 
However, we have made such conservative magnitude and flux cuts to assure the completeness of the sample.
With these ranges,
the number density indeed increases monotonically with decreasing the
flux or increasing the magnitude (see
\cite{Hayashi:2018,Hayashi:2020}) 
and the mean completeness of our [OII] sample becomes $0.974$ for {\it NB816}
and $0.956$ for {\it NB921}.
The numbers of [OII] emitters are
then reduced to 8302 and 9578 for {\it NB816} and {\it NB921} data,
respectively.  See table \ref{tab:four_fields} for the numbers of the
sample for each field.  The line flux cut corresponds to the observed
line luminosity thresholds of $L \geq 2.56\times 10^{41} \rm{[erg~s^{-1}]}$
($z=1.19$) and $L \geq 4.30 \times 10^{41} \rm{[erg ~s^{-1}]}$
($z=1.47$).  The median luminosity with the $2.5$th--$97.5$th percentiles
is $\overline{L} = 4.49^{+12.67}_{-1.82} \times 10^{41}
\rm{[erg ~s^{-1}]}$ at $z=1.19$ and $\overline{L} = 7.88^{+16.95}_{-3.35} \times 10^{41}
\rm{[erg ~s^{-1}]}$ at $z=1.47$. 
Stellar masses for the [OII] emitters are estimated by
spectral energy distribution fit with five HSC BB data in \citet{Hayashi:2020}. 
The median stellar mass of our [OII] emitter sample with the $2.5$th--$97.5$th percentiles
is $\overline{M}_* = 8.58^{+132.54}_{-7.13}\times 10^{9}M_\odot$ 
at $z=1.19$ and $\overline{M}_* = 1.72^{+12.40}_{-1.41}\times 10^{10}M_\odot$ 
at $z=1.47$.

Not all the galaxies in the catalog are real [OII] emitters, but some
are fake lines due to noise and some others are real emission lines
but not [OII] emitters.
The contamination rate is investigated by applying the photo-$z$ and 
color selections to galaxies confirmed with spectroscopic redshifts.
The fraction of such non-[OII] emitters has
been significantly reduced in the current sample compared to that in
the PDR1 catalog because (1) the method of selecting emission-line galaxies has been improved from using the \texttt{cmodel} magnitude to the fixed aperture magnitude \citep{Hayashi:2020} and (2)
the pipeline used for the data process, \texttt{hscPipe}, has been upgraded in the PDR2
\citep{Bosch:2018,Aihara:2019}.  The analysis by \citet{Hayashi:2020} estimated the
fraction of non-[OII] emitters in our sample is less than $10\%$ for
most cases and $\sim 20\%$ at most, by applying photometric redshift and color selections to galaxies confirmed with spectroscopic redshifts. This contaminant fraction can be
further reduced by a machine-learning based technique for photometric
redshift estimates (see e.g., \cite{Hsieh:2014}).  However, since the
current contaminant fraction is low enough for our analysis, we do not
include this process. The angular distribution of the [OII] emitters is shown in
figure~\ref{fig:angular_dist}.  Table \ref{tab:four_fields} summarizes
the properties of the data.

\subsection{Random catalog}

The HSC-SSP PDR2 provides us with the catalog of random points across
the survey area as one of the value-added products (see subsection 5.4
of \cite{Aihara:2019}). The number density of the random points is 100 
per square arcmin. We apply the same basic flags as those for
the emission-line galaxies except for the flags related to the source
detection (i.e., merge\_peak, sdssshape\_flag, cmodel, and psfflux) to
select the random points (see subsection 2.1.1 of
\cite{Hayashi:2020}). We also apply the same bright star masks as
those used for the selection of the emission-line galaxies to the
random points (see \cite{Hayashi:2020} for the details of the masks
and also subsection 6.6.2 of \cite{Aihara:2019}, as well as
\cite{Coupon:2018}). Consequently, the random points used in this study are
distributed over the regions identical to where the emission-line
galaxies are surveyed.

\section{Angular correlation function}\label{sec:measurement}
In this paper we analyze the spatial distribution of [OII] emitters
using the angular correlation function.  The angular correlation
function is related to the spatial correlation function in three
dimension through the relation \citep{Peebles:1973,Peebles:1980}:
\be
w(\theta) = \int dx_1 p(x_1) \int dx_2 p(x_2) \xi(r), \label{eq:3d_to_angular}
\ee
where $r$ is the 3-d separation between two galaxies, $r =
[x_1^2+x_2^2-2x_1x_2\cos{\theta}]^{1/2}$, $x_i=x(z_i)$ is the comoving
distance to a galaxy at redshift $z_i$, and $p(x)$ is the radial
selection function normalized as $\int^\infty_0dxp(x)=1$.

\subsection{Correlation function estimation}
We first measure the angular correlation function of [OII] emitters at
$z=1.19$ and $z=1.47$ from the four fields, COSMOS, DEEP2-3, ELAIS-N1,
and SXDS+XMM-LSS.  We then combine the four correlation functions at
each redshift for the statistical analysis.

We adopt the Landy-Szalay estimator \citep{Landy:1993} to measure the
angular correlation function,
\be
\hat{w}_k(\theta)=\frac{DD_k-2DR_k-RR_k}{RR_k},
\ee
where $DD_k$, $RR_k$, and $DR_k$ are respectively the normalized
counts of data-data, random-random, and data-random pairs at given
angular separation $\theta$. 
The random catalogs contain large numbers of points, which are more than 600 times as dense as our [OII] emitter catalogs to ensure that the shot noise due to the finite random points is negligible.
 The subscript $k$ denotes the $k$th
field $(1\leq k \leq 4)$ and hat means an estimated quantity which can
be different from the true one [see equation (\ref{eq:w_with_ic})
below].  To combine the correlation function measurements from the
four fields, we weigh each measurement by a function $W_k$,
\be
1+\hat{w}(\theta) = \frac{\sum_{k=1}^4W_k^2(\theta) \left( 1+\hat{w}_k(\theta)\right) }{\sum_{k=1}^4W_k^2(\theta)}.
\ee
Thus the combined correlation function is expressed as
\be
\hat{w}(\theta) = \frac{\sum_{k=1}^4W_k^2(\theta) \hat{w}_k(\theta)}{\sum_{k=1}^4W_k^2(\theta)}. \label{eq:w_comb} 
\ee
We adopt the inverse-variance weighting,
$W_k(\theta)=\sigma_k^{-1}(\theta)$, where $\sigma_k^2(\theta)$ is the
diagonal component of the covariance matrix in the $k$-th field (see
next subsection).

\begin{figure}
\begin{center}
\vspace{-.3cm}
\FigureFile(80mm,80mm){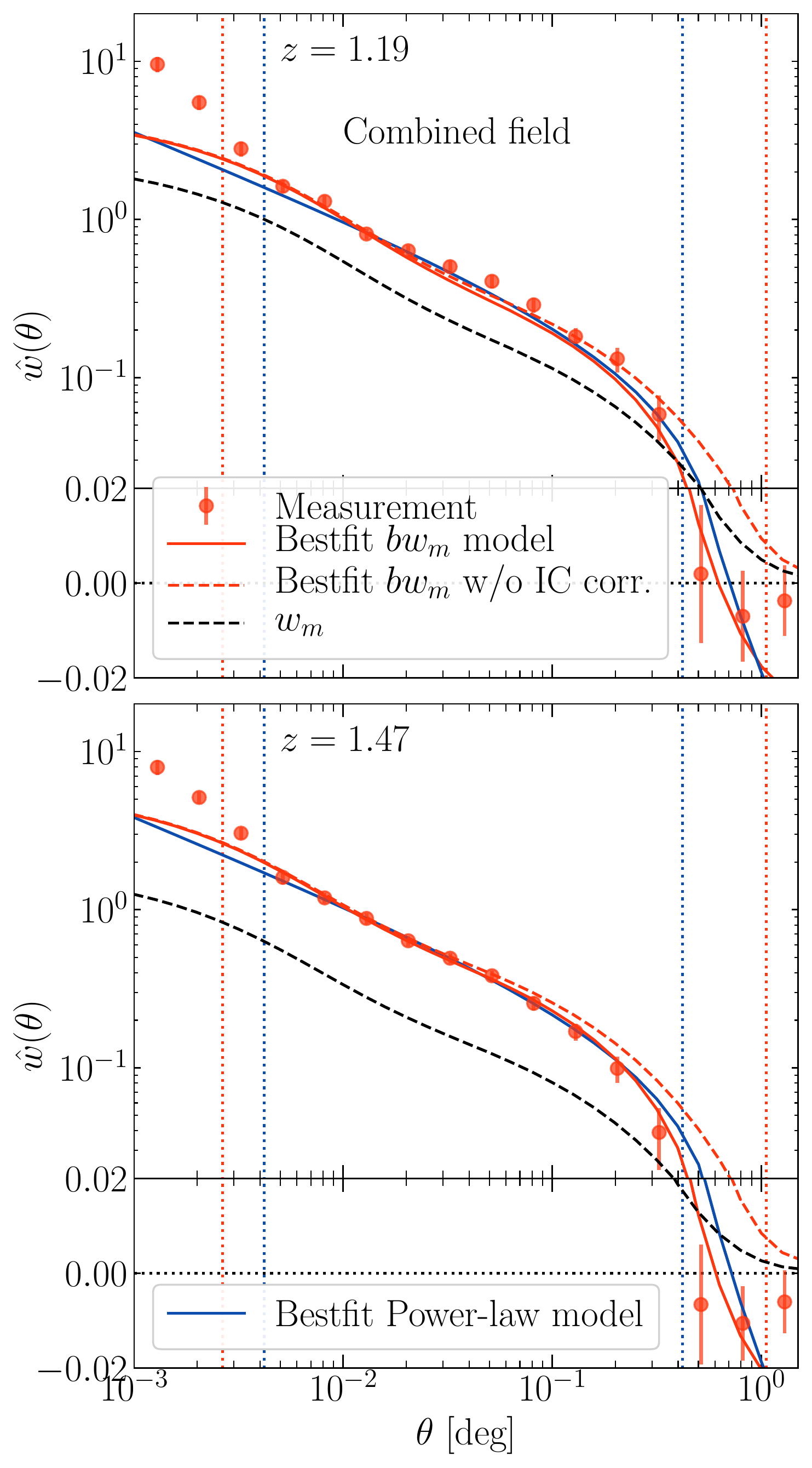}
\end{center}
\caption{Angular correlation functions of [OII] emitters measured from
  the entire four fields at $z=1.19$ (top) and $z=1.47$ (bottom). Note
  that the vertical axis mixes logarithmic and linear scalings.  The
  red points are the measurement without the integral constraint
  correction, $\hat{w}$.  The errorbars are estimated from jackknife
  resampling.  The blue solid curve is the best-fitting model of the
  power-law model, $(1-f_{\rm
    fake})^2(w(\theta;A_w,\beta)-w_\Omega(A_w,\beta))$, with
  $(A_w,\beta,f_{\rm fake})=(1.10,0.534,0.140)$ for $z=1.19$ and
  $(1.17,0.538,0.139)$ for $z=1.47$.  The data enclosed by the two
  blue vertical lines are used for this analysis.  The red solid curve
  is the best-fitting model of the linearly-biased dark matter model,
  $(1-f_{\rm fake})^2(b^2w_m(\theta)-w_\Omega(b))$, with $(b,f_{\rm
    fake})=(1.60, 0.140)$ for $z=1.19$ and $(2.08,0.140)$ for
  $z=1.47$ when the data enclosed by the two red vertical lines are used.  The red dashed curve is the same model but without
  subtracting the integral constraint denoted by IC in the legend,
  $(1-f_{\rm fake})^2b^2w_m(\theta)$.  The non-linear dark matter
  correlation function, $w_m(\theta)$, is shown by the red dashed
  curve as a reference.  }
\label{fig:w_all}
\end{figure}

The correlation function measured above is underestimated by a
constant due to the finite survey area.  The effect is known as the
integral constraint \citep{Peebles:1976} and calculated as
\be
w_\Omega = \frac{1}{\Omega}\int_\Omega d\Omega_1d\Omega_2 w(\theta), \label{eq:ic}
\ee
where the integral is performed over the solid angle of the survey,
$\Omega$.  The integral constraint is calculated by a given model of
the correlation function $w(\theta;\Theta)$, where $\Theta$ is a set
of model parameters (see sections \ref{sec:result} and \ref{sec:hod}).
Note also that the clustering amplitude is further reduced compared to
the true clustering of [OII] emitters due to the contamination of
non-[OII] emitters and noise lines. As seen in section
\ref{sec:result}, we assume that the correlation of non-[OII] emitters
is negligible.  Then the estimated correlation function is related to
the true correlation function $w$ as
\be
\hat{w}(\theta) =  (1-f_{\rm fake})^2(w(\theta) - w_\Omega), \label{eq:w_with_ic}
\ee
where $f_{\rm fake}$ is the fraction of non-[OII] emitters in our
sample, $0.1 \lesssim f_{\rm fake} \lesssim 0.2$ (see section \ref{sec:data_nb}).

The measured correlation function, combined over the four fields,
$\hat{w}$, is shown as the red points in the upper and lower panels of
Figure \ref{fig:w_all} for $z=1.19$ and $z=1.47$, respectively.  The
correlation function becomes negative at scales, $\theta \sim 1 $ deg ($r\sim 50h^{-1}$Mpc),
due to the integral constraint $w_\Omega$. For the model parameter
fitting below, this effect is properly taken into account.  In this
figure we show various lines, which are the predictions based on the
power-law model and the linearly-biased dark matter model.  We discuss
these in detail in section \ref{sec:result}.
The measurements of the angular correlation functions for individual
fields are presented for a consistency check in appendix
\ref{sec:w_individual}.

\subsection{Covariance matrix}\label{sec:covariance}
We use the Jackknife resampling technique to estimate the covariance
error matrix (see, e.g., \cite{Lupton:1993}).  The covariance of the
angular correlation function for $k$-th field, $C_{k,ij}\equiv C\left(
\hat{w}_k(\theta_i), \hat{w}_k(\theta_j)\right)$, is then estimated as
\bey
  C_{k,ij} &=& 
\frac{N_{\rm sub}-1}{N_{\rm sub}} \nn \\ 
&& \times\sum^{N_{\rm sub}}_{\ell=1}\left(\hat{w}^\ell_k(\theta_i)-\bar{\hat{w}}_k(\theta_i)\right)
\left(\hat{w}^\ell_k(\theta_j)-\bar{\hat{w}}_k(\theta_j)\right),
  \label{eq:cov}
\eey
where $\hat{w}^\ell_k(\theta_i)$ is the correlation function in the
$i$-th angular bin in $\ell$-th jackknifed realization, $N_{\rm sub}$
is the number of realizations in the $k$-th field and
$\bar{\hat{w}}_k(\theta_i)=N_{\rm sub}^{-1}\sum^{N_{\rm
    sub}}_{\ell=1}\hat{w}^\ell_k(\theta_i)$.  The values of $N_{\rm
  sub}$ for each field are shown in table~\ref{tab:four_fields}.  In
total, we use 115 and 119 jackknifed realizations for the $z=1.19$ and
$1.47$ data, respectively. 
We use the $k$-means algorithm
\footnote{ https://github.com/esheldon/kmeans\_radec } to divide our
survey regions to subregions with the same angular areas.
Each subsample includes a region continuous on the sky, $0.38~{\rm deg}$ on a side, which corresponds to the comoving size of $17h^{-1}$Mpc at $z=1.19$ and $20h^{-1}$Mpc at $z=1.47$.
Just like equation (\ref{eq:w_comb}), the covariance matrixes from the four
fields can be combined to $C_{ij}\equiv C\left( w(\theta_i),
w(\theta_j)\right)$ as
\be
C_{ij} = \frac{\sum_{k=1}^4 W_k^2(\theta_i)W_k^2(\theta_j)C_{k,ij}}
{\sum_{k=1}^4W_k^2(\theta_i)\sum_{k=1}^4W_k^2(\theta_j)}~. \label{eq:cov_comb}
\ee
The square roots of the diagonal components of the covariance matrix,
$C_{ii}^{1/2}$, are shown as the error bars in figure \ref{fig:w_all}.

\section{Results}\label{sec:result}
\subsection{Setup}
In this section we consider two simple models for the angular
correlation function, the power-law and linearly biased dark matter
models in sections \ref{sec:power_law} and \ref{sec:nl_matter}, to
model the clustering amplitude of [OII] emitters.  The HOD modeling is
performed in section \ref{sec:hod}.

We adopt the small-angle approximation, known as the Limber's
approximation \citep{Limber:1954,Peebles:1980}.  With this, the
relation between the angular and spatial correlation functions given
by equation (\ref{eq:3d_to_angular}) can be simplified to:
\be
w(\theta) = 2\int^\infty_0 dx p^2(x) \int^\infty_0 dh \xi\left( r=\sqrt{x^2\theta^2+h^2} \right), \label{eq:angular_limber}
\ee
where $x=(x_1+x_2)/2$ and $h=x_2-x_1$.  The validity of the Limber's
approximation for the angular correlation with large angle separations
has been extensively discussed by, e.g., \citet{Simon:2007} and
\citet{Crocce:2011}.  The approximation breaks down on large scales
when the width of the redshift distribution, $z_{\rm min}\leq z \leq
z_{\rm max}$, is too narrow.  The width of our NB filters is $\Delta x
\equiv x_{\rm max} - x_{\rm min} \simeq 50h^{-1}{\rm Mpc}$, where
$x_{\rm min} = x(z_{\rm min})$ and $x_{\rm max} = x(z_{\rm max})$, and
our analysis focuses on the angular scales of $\theta <1~{\rm deg}$.
We thus can safely rely on this approximation.  

Although determination of the true radial selection function is not straightforward, 
it is known to have a form similar with the transmission curve to some extent (see, e.g., figure 3 of \cite{Hayashi:2015}).
However, the
width $\Delta x \simeq 50h^{-1}{\rm Mpc}$ is narrow enough that we can
simply assume the radial selection function to be a constant,
\bey
p(x)= \left\{ \begin{array}{ll}
1/\Delta x& {\rm for }~ x_{\rm min} \leq x \leq x_{\rm max},  \\
0 & {\rm otherwise}.
\end{array}
\right. \label{eq:radial_selection}
\eey
Using the transmission curves available for the NB filters\footnote{https://subarutelescope.org/Observing/Instruments/HSC/sensitivity.html}, we test the effect of choices of $p(z)$ on the calculation of the angular correlation functions. 
We compare the angular correlation function computed using $p(z)$ from the transmission curve with that from the constant $p(z)$ in equation (\ref{eq:radial_selection}). We then find that the correlation functions $w(\theta)$ have exactly the same shape with the amplitude different by $\leq 7\%$ for both {\it NB816} and {\it NB921}. Thus, constraints on the parameters for the correlation function are not affected by the choice of $p(z)$ but it changes only the normalization, $f_{\rm fake}$, by at most $4\%$, much smaller than our $1-\sigma$ prior on $f_{\rm fake}$, $\sim 40\%$ (see the following paragraphs of this subsection). In the following analysis, we thus adopt the simple constant radial selection function [equation (\ref{eq:radial_selection})].

Given a model for the correlation function with a set of parameters
$\Theta$, $w(\theta; \Theta)$, the integral constraint
[equation (\ref{eq:ic})] can be estimated as
\be
w_\Omega(\Theta) = \frac{\sum_i{w(\theta_i;\Theta)RR(\theta_i)}} {\sum_i{RR(\theta_i)}}, \label{eq:ic_rr}
\ee
where $\theta_i$ is the $i$th separation bin and the sum is calculated
over the entire bins.  The normalized random-random pair count for the
combined field, $RR$, is computed in the same way as the correlation
function in equation (\ref{eq:w_comb}),
\be
RR(\theta) = \frac{\sum_{k=1}^4W_k^2(\theta) RR_k(\theta)}{\sum_{k=1}^4W_k^2(\theta)}. \label{eq:rr_comb}
\ee

As discussed in section \ref{sec:data_nb}, not all the lines detected
as [OII] emitters are actually [OII] lines, but some are other
emission lines or just noises.  The fraction of fake [OII] lines is
denoted as $f_{\rm fake}$.  Then the observed correlation function has
contributions not only from the auto-correlation of [OII] emitters but
also from the auto-correlation of fake lines and their
cross-correlation, which are suppressed by factors of $f_{\rm fake}^2$
and $2f_{\rm fake}(1-f_{\rm fake})$, respectively (see e.g.,
\cite{van_den_Bosch:2013,Okumura:2015,Okumura:2017} for detailed
theoretical schemes for the tracer decomposition).  The contamination
fraction is estimated to be around $f_{\rm fake}\sim 0.14$ by applying
photometric redshift and color selections to galaxies confirmed with
spectroscopic redshifts \citep{Hayashi:2020}.  Thus, even though the
non-[OII] emitters have a non-negligible auto-correlation, the
(intrinsically small) clustering amplitude is suppressed by $0.14^2
\sim 0.02$.  We therefore can safely assume that the observed
correlation is dominated by the auto correlation of [OII] emitters
(see, e.g., \cite{Okumura:2016,Kashino:2017a} for a similar
treatment)\footnote{If $f_{\rm fake}$ becomes large so that the
contribution from non-[OII] emitters is non-negligible, a
sophisticated treatment of the line contaminant for the cosmological
analysis is required (e.g., \cite{Leung:2017,Addison:2019}).}.  The
amplitude of the correlation function is thus simply scaled by a
factor of $(1-f_{\rm fake})^2$.  We treat $f_{\rm fake}$ as a free
parameter with a prior of $f_{\rm fake}=0.140\pm 0.060$ to cover 
the uncertainties described in section \ref{sec:data_nb} within the $1-\sigma$ confidence level.
We chose this uncertainty to be very conservative and much larger than 
it is to avoid inducing biased constraints on model parameters by imposing 
strong (and possibly incorrect) priors. 
The imposed prior is coincidentally equivalent to that adopted in \citet{Kashino:2017a}.

Finally, once the model with parameters for the correlation function
is given as $w(\theta;\Theta)$, the theoretical prediction to be
compared to the observed correlation function, $\hat{w}$, is obtained
with an additional parameter $f_{\rm fake}$ through equation
(\ref{eq:w_with_ic}).  To constrain model parameters, the $\chi^2$
statistic is calculated for the correlation function, $w$, as
\be
\chi_w^2 (\Theta)= \sum_{i=1}^{N_{\rm bin}}\sum_{j=1}^{N_{\rm bin}} {\Delta_i C^{-1}_{ij} \Delta_j}, \label{eq:chi2_w}
\ee
where $N_{\rm bin}$ is the number of angular separation bins used in
the analysis, $\Delta_i = \hat{w}_{\rm obs}(\theta_i)-\hat{w}_{\rm
  th}(\theta_i;\Theta)$ is the difference between the observed
correlation function [equation~(\ref{eq:w_comb})] and the theoretical
prediction [equation (\ref{eq:w_with_ic})] with $\Theta$ being a set of
parameters to be constrained, and $C^{-1}_{ij}$ is the inverse
covariance matrix [equation (\ref{eq:cov_comb})] multiplied by a
correction for the finite realization effect, $(N_{\rm sub}-N_{\rm
  bin}-2)(N_{\rm sub}-1)^{-1}$ \citep{Hartlap:2007}.

In order to perform a maximum likelihood analysis, we use the Markov
chain Monte Carlo (MCMC) sampler \texttt{emcee}
\citep{Foreman-Mackey:2013}.  

\subsection{Power-law constraints} \label{sec:power_law}
It has been known that shapes of the spatial and angular correlation
functions approximately follow a power law, originally found by
\citet{Totsuji:1969} (see also \cite{Peebles:1980}), as
\be
w(\theta; A_w, \beta) = A_w \left(
\frac{\theta}{1~{\rm arcmin}}
\right)^{-\beta}.
\ee
The $\chi^2$ statistic is calculated with $\Theta=(A_w,\beta,f_{\rm
  fake})$.  Because a power-law model is known to fail to fit an
observed correlation function at small- and large-scale limits, we
compute the $\chi^2$ statistic for a relatively narrow angular
separation range, $0.004 < \theta< 0.4~[{\rm deg}]$, denoted by the
blue vertical lines in figure \ref{fig:w_all}. Thus the number of bins
is $N_{\rm bin}=10$.

\begin{figure}[bt]
\begin{center}
\vspace{-.5cm}
\FigureFile(78mm,78mm){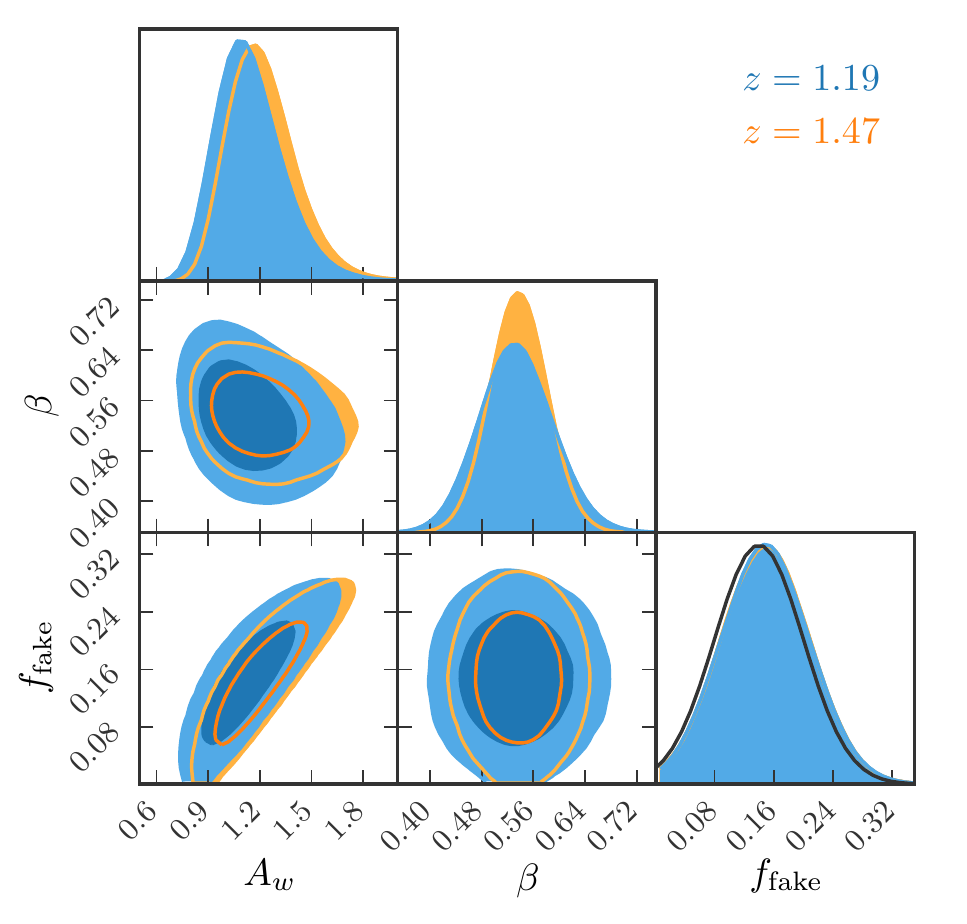}
\end{center}
\caption{Constraints on the power-law model parameters $(A_w,\beta)$
  and the fake line fraction parameter $f_{\rm fake}$ from the angular
  clustering of [OII] emitters for $z=1.19$ (blue) and $z=1.47$
  (orange). Contours show the $68\%$ and $95\%$ confidence levels. The
  diagonal panel show the posterior probability distribution of each
  parameter. A Gaussian prior is adopted for the fake line fraction,
  $f_{\rm fake}=0.140\pm 0.060$, which is shown as the black curve compared with the one-dimensional
  posterior. The posterior of $f_{\rm fake}$ for $z=1.47$ (orange) is almost entirely behind the one for $z=1.19$. } 
\label{fig:result_powerlaw}
\end{figure}

\begin{table*}[bt!]
\caption{Parameter constraints on single power-law and linearly-biased dark matter models${}^*$ }
\begin{center}
\begin{tabular}{ll | ll | ll}
\hline 
& & \multicolumn{2}{c|}{{\it NB816}}  &  \multicolumn{2}{c}{{\it NB921}}   \\ 
Power-law parameter  & Prior & Best-fit & Posterior PDF & Best-fit & Posterior PDF \\
\hline 
$A_w$ (at $1$ arcmin)   & None & $1.10$ &   $1.12^{ + 0.21} _{- 0.16}$   & $1.17$ & $1.20^{ + 0.20}_{ - 0.17}$ \\ 
$\beta$                    & None        & $0.534$ & $0.533 ^{+ 0.057}_{ - 0.055}$ & $ 0.538$ & $0.537^{+ 0.044}_{ - 0.043}$  \\
$f_{\rm fake}$& $0.140\pm0.060$ & $0.140$ & $0.149^{ + 0.059}_{ - 0.059}$ & 0.139 & $0.149^{ + 0.060}_{ - 0.059}$  \\
$r_0(h^{-1}{\rm Mpc})$ & $-$       & $4.08$ & $4.12^{ + 0.50}_{ - 0.41}$ & $ 4.55$ & $4.61^{+ 0.51}_{- 0.45}$  \\
\hline 
Linearly-biased DM Parameter  & Prior & Best-fit & Posterior PDF & Best-fit & Posterior PDF \\
\hline
$b$                 & None & $1.60$ & $1.61 ^{+ 0.13}_{ - 0.11}$ & 2.08 & $2.09^{ + 0.17}_{ - 0.15}$  \\
$f_{\rm fake}$& $0.140\pm0.060$ & $0.140$ & $0.145^{ + 0.059}_{ - 0.059}$ & 0.140 & $0.145^{ + 0.060}_{ - 0.059}$  \\
\hline
\end{tabular}
\end{center}
\label{tab:powerlaw_nl}
\begin{tabnote}
$*$ The values in the column of ``Prior'' quote the mean and standard
  deviation of the Gaussian priors. The column of ``Best-fit'' shows
  the parameter set which gives the minimum value of $\chi^2$. In the
  column of ``Posterior PDF'', the central value is a median and the
  error means $16-84$ percentiles after other parameters are
  marginalized over.
\end{tabnote}
\end{table*}

Figure \ref{fig:result_powerlaw} shows the joint constraints on the
power-law parameters and the fake line fraction, $(A_w, \beta, f_{\rm
  fake})$ with blue and orange contours for $z=1.19$ and $z=1.47$,
respectively\footnote{To show parameter constraints using the
MCMC sampling here and in the rest of this paper, we use a python package \texttt{pygtc}
\citep{Bocquet:2016}.}.  
The parameter set with the minimum value of $\chi^2$
for $z=1.19$ is $(A_w, \beta, f_{\rm fake})=(1.10, 0.534, 0.140)$
while that for $z=1.47$ is $(A_w, \beta, f_{\rm fake})= (1.17, 0.538,
0.139)$.  With these sets of the best-fitting parameters, we derive
the integral constraints as $w_\Omega = 0.110$ and $w_\Omega = 0.114$,
for lower and higher redshifts, respectively.  The best-fitting
power-law model prediction for the measured correlation function,
$(1-f_{\rm fake})^2(A_w\theta^{-\beta}-w_\Omega)$, is shown by the
blue line in figure \ref{fig:w_all}.  The constrained parameters are
shown in table \ref{tab:powerlaw_nl}.

When the two parameters for a power-law model of the angular
correlation, $(A_w,\beta)$, are known, one can determine its
3-dimensional clustering, $\xi(r) = (r/r_0)^{-\gamma}$, as
\citep{Peebles:1980,Efstathiou:1991,Simon:2007}
\bey
A_w &=& r_0^\gamma \left( \frac{10800}{\pi} \right)^{\gamma-1} 
B\left( \frac{1}{2},\frac{\gamma-1}{2} \right)\int^\infty_0 dx p^2(x) x^{1-\gamma} \nn \\ 
&=&r_0^\gamma  
\frac{10800^{\gamma-1}\left(x_{\rm max}^{\gamma-1}-x_{\rm min}^{\gamma-1}\right)}{\pi^{\gamma-1}(2-\gamma)(x_{\rm max}-x_{\rm min})^2}
B\left( \frac{1}{2},\frac{\gamma-1}{2} \right), \\
\beta &=& \gamma -1,
\eey
where $B$ is the beta function.  The final expression of $A_w$ is
derived using the constant radial selection function [equation
  (\ref{eq:radial_selection})].  As a result, the clustering length is
constrained to be $r_0= 4.12^{+0.50}_{-0.41}$ at $z=1.19$ and $r_0=
4.61^{+0.51}_{-0.45}$ at $z=1.47$. These values correspond to the host
halo masses of $M\gtrsim 10^{12} M_\odot /h$ and are roughly
consistent with the preceding results of ELG clustering (see e.g.,
\cite{Kashino:2017a}).  Here we do not make a detailed comparison with
previous works nor argue the clustering evolution further because the
power-law model is not so accurate to be used to interpret the precise
correlation function measurement anyways.  In section \ref{sec:hod},
we discuss physical properties of halos hosting ELGs based on an HOD
model that is much more sophisticated than the power-law model.

\subsection{Linearly biased dark matter model}\label{sec:nl_matter}
Next, we consider another simple model, the non-linear correlation
function with the linear galaxy bias factor \citep{Kaiser:1984},
\be
\xi(r;b)=b^2\xi_m(r), \label{eq:nl_matter3d}
\ee
where $\xi_m$ is the underlying matter correlation function.  We use
the fitting formula of \citet{Takahashi:2012} to compute the nonlinear
matter power spectrum, $P_m(k)$, and then Fourier transform it to obtain $\xi_m$.
Given equation (\ref{eq:nl_matter3d}), a similar form is derived for
the angular correlation function as
\be
w(\theta;b)=b^2w_m(\theta), \label{eq:nl_matter}
\ee
where $w_m$ is the angular correlation function of matter, calculated
using equation (\ref{eq:angular_limber}) with $\xi(r)$ replaced by the
3-dimensional mater correlation function, $\xi_m(r)$.  In this model
there are two free parameters to be determined, $\Theta=(b,f_{\rm
  fake})$.  This model is valid for a wider range of angular scales
than the single power-law model.  The data at the scales $0.0025 <
\theta < 1~[{\rm deg}]$ are used to calculate the $\chi^2$ statistic
($N_{\rm bin}=13$).

\begin{figure}[bt]
\begin{center}
\vspace{-.5cm}
\FigureFile(58mm,58mm){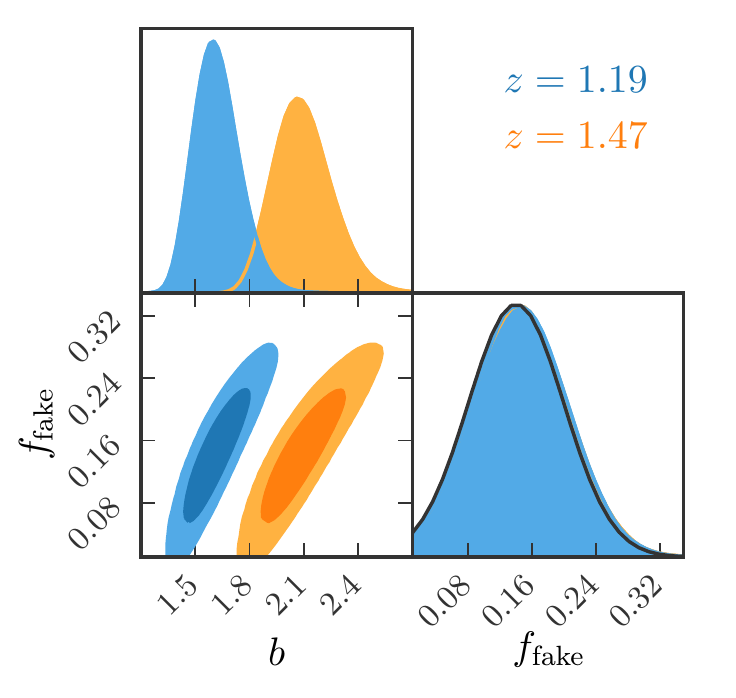}
\end{center}
\caption{Constraints on the linear galaxy bias $b$ of the nonlinear
  dark matter model and the fake line fraction parameter $f_{\rm
    fake}$ for $z=1.19$ (blue) and $z=1.47$ (orange). Contours show
  the $68\%$ and $95\%$ confidence levels. The diagonal panel show the
  posterior probability distribution of each parameter. A Gaussian
  prior is adopted for the fake line fraction, $f_{\rm fake}=0.140\pm
  0.060$, which is shown as the black curve compared with the one-dimensional
  posterior. The posterior of $f_{\rm fake}$ for $z=1.47$ (orange) is almost entirely behind the one for $z=1.19$.
  }
\label{fig:result_nlmatter}
\end{figure}

The joint constraints on $(b,f_{\rm fake})$ are shown in figure
\ref{fig:result_nlmatter}.  We find the best-fitting value of the bias
parameter, $b= 1.61 ^{+ 0.13}_{ - 0.11}$ and $b= 2.09^{ + 0.17}_{ -
  0.15}$, respectively for the lower and higher redshift measurements. 
These bias values can be compared to the effective bias $b_{\rm eff}$
to be determined by a HOD modeling [equation (\ref{eq:b_hod})].  The
integral constraint is calculated by these best-fitting parameter sets
for the correlation in the combined field as $w_\Omega = 0.0270$ and
$w_\Omega = 0.0277$ for $z=1.19$ and $1.47$, respectively.

The best-fitting linearly-biased dark matter model, $(1-f_{\rm
  fake})^2(b^2w_m(\theta)-w_\Omega)$, is shown as the red solid curve
in figure \ref{fig:w_all}.  To see the contribution of the integral
constraint, we show the same model without $w_\Omega$ term, $(1-f_{\rm
  fake})^2b^2w_m(\theta)$, as the red dashed curve. As expected, the
model correlation function is positive over all the scales studied
here.
As a reference, the dark matter correlation function, $w_m$, is shown
by the black dashed curve in figure \ref{fig:w_all}.

The relatively large uncertainties on the bias parameter comes from 
our conservative prior on $f_{\rm fake}$. 
To see this, let us directly constrain the observed amplitude by  
considering the relation:
\bey
\hat{w}(\theta;\wt{b}) &=& \wt{w}(\theta;\wt{b})-\wt{w}_\Omega(\wt{b}) \nn \\
 &=& \wt{w}(\theta;\wt{b}) - \frac{\sum_i{\wt{w}(\theta_i;\wt{b})RR(\theta_i)}} {\sum_i{RR(\theta_i)}}, \label{eq:nl_matter_wt}
\eey
where $\wt{b}$ is the bias parameter of the observed field, $\wt{w}(\theta,\wt{b}) = \wt{b}^2w_m(\theta)$. 
Equation (\ref{eq:nl_matter_wt}) becomes equivalent to equations (\ref{eq:w_with_ic}) and (\ref{eq:nl_matter}) 
if we set $\wt{b} \equiv (1-f_{\rm fake})b$.
By adopting this 1-parameter linearly-biased dark matter model, we obtain
tighter constraints as $\wt{b} = 1.379^{+0.042}_{-0.043}$ and $\wt{b} = 1.786 ^{+0.052}_{-0.054}$ at $z=1.19$ and $z=1.47$, respectively, without any prior.

In the next subsection we further analyze the [OII] emitter samples with different luminosity and stellar mass thresholds using the linearly-biased DM model.

\begin{figure}[bt]
\begin{center}
\vspace{-.2cm}
\FigureFile(80mm,80mm){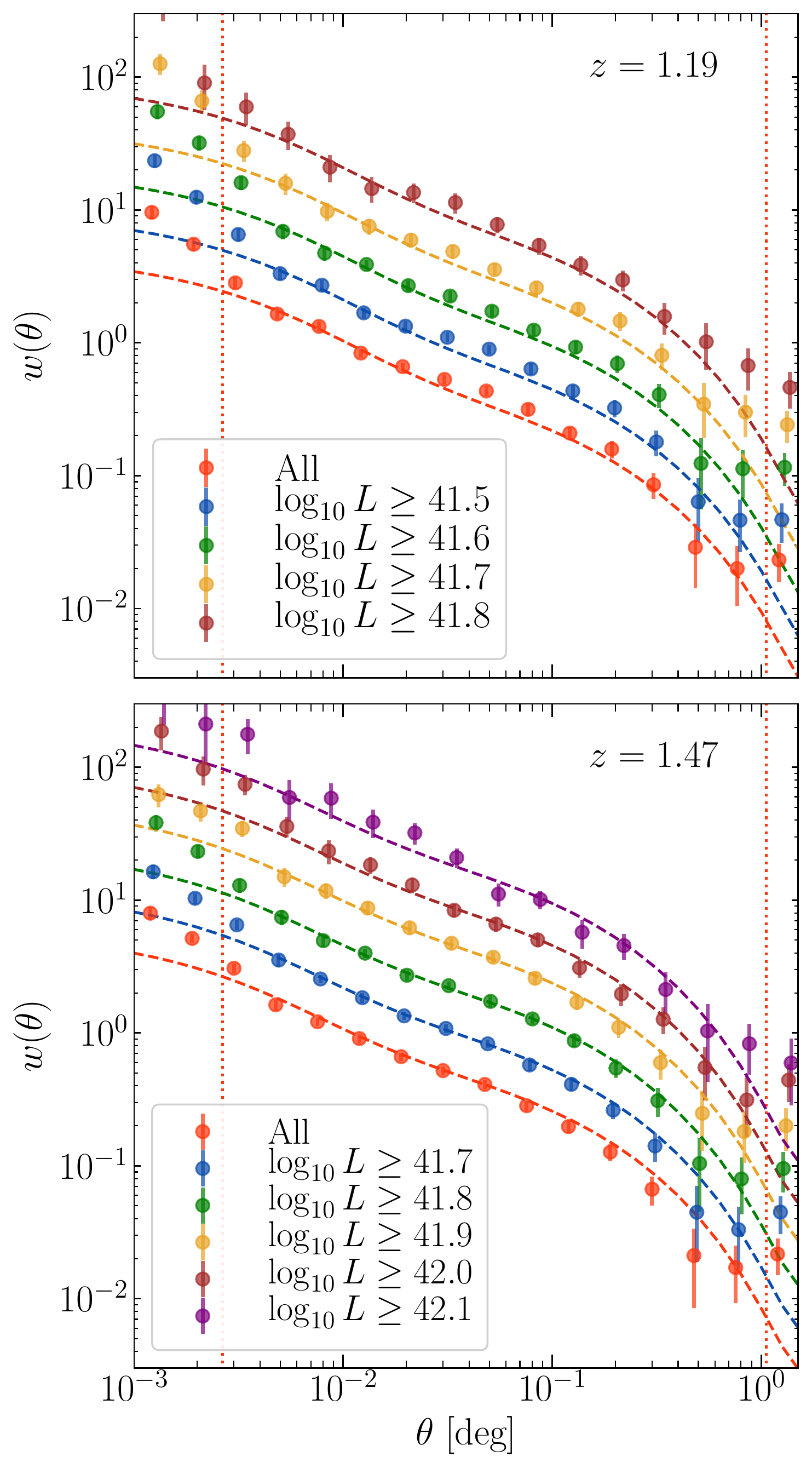}
\end{center}
\caption{
Angular correlation functions of [OII] emitters at $z=1.19$ (top) and $z=1.47$ (bottom) with different peak luminosity cuts. The unit of $L$ in the legend is $[{\rm erg~s}^{-1}]$. For the visualization purpose, the integral constraint is taken into account, and thus the vertical axis presents $\hat{w}+(1-f_{\rm fake})^2w_\Omega$. The amplitudes of the correlation functions shown by the blue, green, yellow, brown and purple points are multiplied by $2^n$ where $n=1,2,\cdots ,5$, respectively. The dashed curves are the best-fitting models of the linearly-biased dark matter model, $(1-f_{\rm fake})^2b^2 w_m(\theta)$.}
\label{fig:w_all_dlt_lum}
\end{figure}

\begin{figure}[bt]
\begin{center}
\vspace{-.2cm}
\FigureFile(78mm,78mm){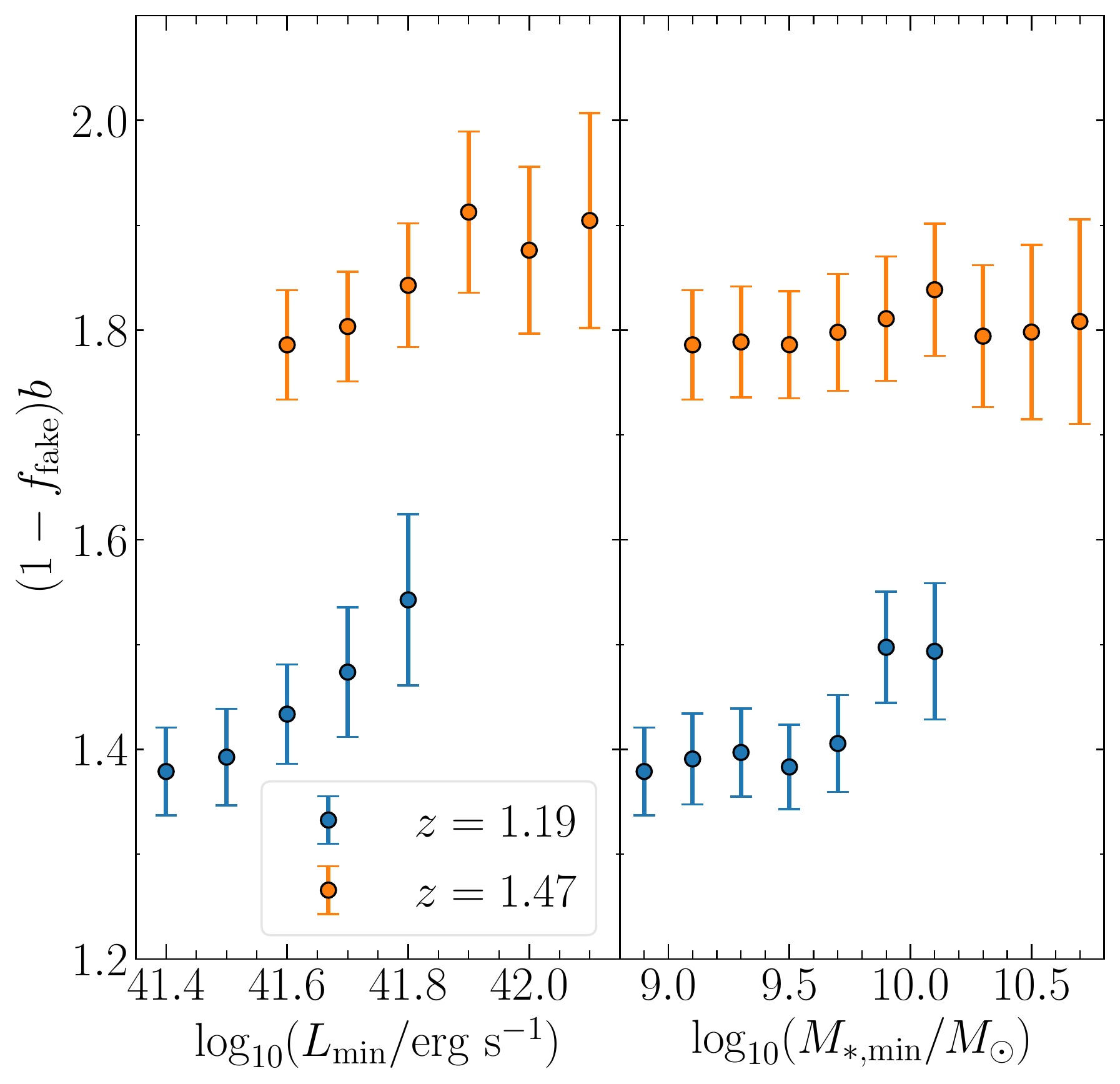}
\end{center}
\caption{
Constraints on the measured clustering amplitude of [OII] emitters, $\wt{b}=(1-f_{\rm fake})b$, as a function of the minimum peak luminosity $L_{\rm min}$ (left) and stellar mass $M_{*,{\rm min}}$ (right). The blue and orange points are the results at $z=1.19$ and $z=1.47$, respectively.
}
\label{fig:bias_lum}
\end{figure}

\subsection{Line luminosity and stellar mass dependences of [OII]-emitter clustering}\label{sec:bias_lum}
It is interesting to see how the amplitude of a measured correlation function
depends on properties of  the [OII] emitters. 
Here we consider two basic properties, the luminosity of the [OII] emission line and the stellar mass of the 
host galaxies, and split our sample into subsamples based on these thresholds.

Figure \ref{fig:w_all_dlt_lum} shows the angular correlation functions 
measured with different line luminosity cuts. 
For clarity, we add the integral constraint correction so that all the correlation functions 
become positive at all scales probed. 
For each correlation function, we perform a model fitting using the linearly biased dark matter model  
considered in section \ref{sec:nl_matter}.
The best-fitting model for each measurement, $(1-f_{\rm fake})^2b^2w_m(\theta)$, 
is presented by the dashed curve. 
As discussed in section \ref{sec:nl_matter}, this simple model fails to explain the measured 
correlation function at small scales. 
To see the dependence of the clustering amplitude on the luminosity, 
we constrain the parameter $\wt{b}$ in equation (\ref{eq:nl_matter_wt}) as a function of the minimum line luminosity $L_{\rm min}$.
The result is shown in the left panel of figure \ref{fig:bias_lum}.
For each redshift, we find that the bias $b$ increases with increasing line luminosity, by $\sim 10\%$. 
If the fake line fraction, $f_{\rm fake}$, changed significantly with the luminosity, this luminosity dependence would be modulated. 

We repeat the same analysis by splitting our [OII] emitter sample into subsamples with 
stellar mass cuts $M_{*,{\rm min}}$. We measure the angular correlation function of [OII] emitters with the stellar mass of $M_* \geq M_{*,{\rm min}}$.
Since the fitting result for $w(\theta)$ is similar to that with the luminosity cuts in figure \ref{fig:w_all_dlt_lum}, 
we do not plot the figure. 
We constrain the measured clustering amplitude $\wt{b}$ as a function of $M_{*,{\rm min}}$, as shown in the right panel of figure \ref{fig:bias_lum}. 
A trend similar to the line luminosity limited samples is found: 
[OII] emitters with higher stellar masses cluster more strongly than those with lower masses. 
Almost no mass dependence is seen in the clustering at $z=1.47$. 
As discussed in section 3.6 of \citet{Hayashi:2018}, 
multi-band data at longer wavelength, such as near-infrared data, would be 
required to accurately estimate the stellar masses at $z\sim 1.5$. 
Thus, our result for the stellar mass dependence of the clustering at $z=1.47$ needs to be interpreted with caution. 
Overall, however, the dependence of clustering of [OII] emitters on stellar masses is not as strong as 
that on line luminosities, 
which is qualitatively consistent with the result of \citet{Khostovan:2018}. 

\section{Dark matter-[OII] emitter connections} \label{sec:hod}
The halo occupation distribution (HOD) modeling is an empirical and
parametric way to describe the connection between galaxies and their
host halos through the modeling of the abundance and clustering of
galaxies
\citep{Jing:1998a,Peacock:2000,Seljak:2000,Ma:2000,Scoccimarro:2001,Cooray:2002,Berlind:2002,Zheng:2005,van_den_Bosch:2013,Hikage:2013a,Okumura:2015}.
In this section we use HOD modeling to investigate physical properties
of halos which host [OII] emitters.

\subsection{Formalism of halo model}\label{sec:halo_model}
The mean number of galaxies residing in a halo of mass $M$, denoted as
$\left\langle N(M) \right\rangle$, is decomposed into central and
satellite components,
\begin{eqnarray}
  && \left\langle N(M) \right\rangle =  \left\langle N_{\rm cen}(M)\right\rangle + \left\langle N_{\rm sat}(M)\right\rangle.
\end{eqnarray}
Once the HOD model is specified, the average number density of
galaxies is calculated as
\be
  n_g=\int \left\langle N(M) \right\rangle n(M)dM, \label{eq:ng_hod}
\ee
where $n(M)dM$ is the halo mass function.

In a halo model approach, the power spectrum of galaxies, $P(k)$ where
$k$ is the wavenumber, can be decomposed into the one- and two-halo
terms,
\be
P(k)=P_{1h}(k)+P_{2h}(k).
\ee
The one-halo term can be further decomposed into contributions from
the clustering of central-satellite and satellite-satellite pairs
hosted by the same halos, namely,
\bey
P_{1h}(k) &=& \frac{1}{n_g^2}\int n(M)dM \left[ 2\left\langle N_{\rm cen}N_{\rm sat} \right\rangle u(k,M) \right. \nn \\ 
&& \ \ \ \ \ \ \ \ \ \ \ \ \ \ \ \ \ \ \ \ \ \ \ \ \ \ \  \left. + \left\langle N_{\rm sat}(N_{\rm sat}-1) \right\rangle u ^2 (k,M)\right],
\eey
where $u(k,M)$ describes the Fourier transform of the density profile
of satellite galaxies within dark matter halos, which is assumed to
follow the matter density profile. We assume that $N_{\rm sat}$
follows the Poisson statistics such that $\langle N_{\rm sat}(N_{\rm
  sat}-1) \rangle = \langle N_{\rm sat}\rangle^2$.  We also assume
that the occupation numbers of central and satellite galaxies are
independent, i.e., $\langle N_{\rm cen} N_{\rm sat}\rangle = \langle
N_{\rm cen}\rangle \langle N_{\rm sat}\rangle$.  The two-halo term can
be similarly decomposed into central-central, central-satellite and
satellite-satellite galaxy pairs hosted by distinct halos,
\bey
P_{2h}(k) &=& \frac{1}{n_g^2}\int n(M)dM \int n(M')dM' \nn \\
&&\ \ \ \times \left [ \left \langle N_{\rm cen} \right \rangle + \left \langle N_{\rm sat}\right \rangle u(k,M)\right ] \nn \\
&&\ \ \ \times \left [ \left \langle N_{\rm cen} \right \rangle + \left \langle N_{\rm sat}\right \rangle u(k,M')\right ]
P_{hh}(k;M,M'),
\eey
where $P_{hh}(k;M,M')$ is the cross-power spectrum of halos with
masses $M$ and $M'$. 
The halo mass function is calculated using the
model derived by \citet{Tinker:2010}.  The density profile of halos is
assumed to have the NFW profile \citep{Navarro:1997} with the
concentration given by \citet{Duffy:2008}.
We compute $P_{hh}$ as a product of the nonlinear matter power spectrum \citep{Takahashi:2012} and the halo bias,
\be
P_{hh}(k;M,M') = b_h(M)b_h(M')P_m(k),
\ee
where $b_h(M)$ is the bias for a halo of mass $M$.
We use the large-scale halo bias $b_h(M)$ proposed also by \citet{Tinker:2010}.

The three-dimensional correlation function of galaxies is calculated
by a Fourier transform of the power spectrum,
\be \xi(r) = \frac{1}{2\pi^2} \int^\infty _0 dkk^2P(k) \frac{\sin{kr}}{kr}. \ee
Then the angular correlation function is finally obtained through
equation (\ref{eq:angular_limber}).

\begin{figure*}[bt]
\begin{center}
\vspace{-1.cm}
\FigureFile(175mm,175mm){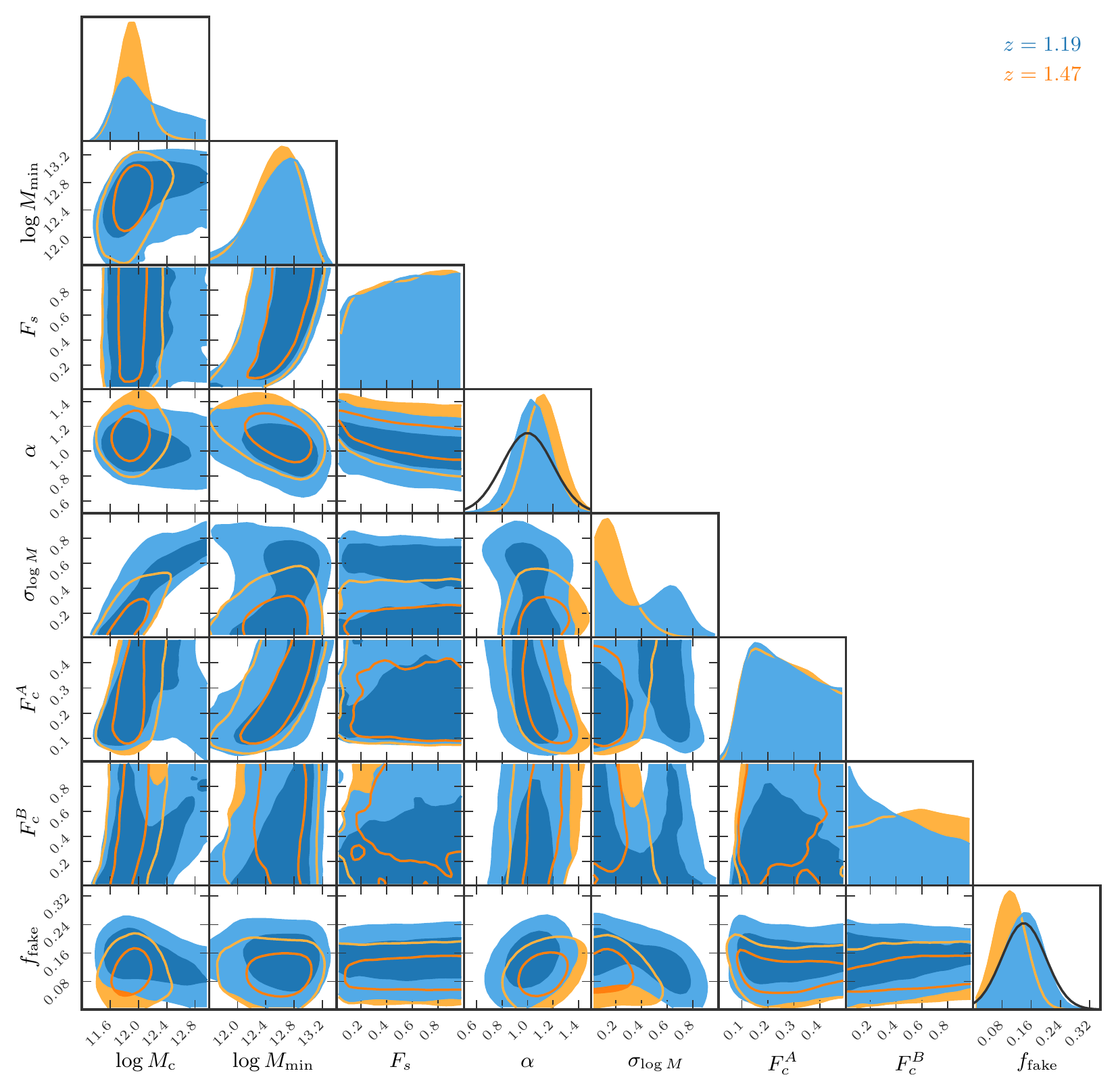}
\end{center}
\caption{Constraints on parameters of Geach HOD model, $(\log{M_{\rm
      c}}, \log{M_{\rm min}}, \alpha, \sigma_{\log{M}}, F_s,
  F_c^A,F_c^B) $ and $f_{\rm fake}$ for $z=1.19$ (blue) and $z=1.47$
  (orange). Two-dimensional contours show the $68\%$ and $95\%$
  confidence levels after the other six parameters marginalized
  over. The diagonal panels show the posterior probability distribution
  of each parameter. Gaussian priors are assumed for $\alpha$ and
  $f_{\rm fake}$, as depicted by the black sold curves in the 1-d
  posterior panels.  }
\label{fig:result_hod_contreras}
\end{figure*}

In addition to the number density, given a set of HOD parameters, one
can determine various physical quantities such as the effective bias,
\be
b_{\rm eff} = \frac{1}{n_g}\int b_h(M) \left\langle N(M) \right\rangle n(M)dM, \label{eq:b_hod}
\ee
the effective halo mass,
\be
M_{\rm eff} = \frac{1}{n_g}\int M\left\langle N(M) \right\rangle n(M) dM, \label{eq:m_hod}
\ee
and the satellite fraction,
\be
f_{\rm sat} = \frac{1}{n_g} \int \left\langle N_{\rm sat}(M) \right\rangle n(M) dM. \label{eq:fsat_hod}
\ee
%

\begin{table*}[bt!]
\caption{Priors and constraints of the HOD parameters for Geach model${}^*$
}
\begin{center}
\begin{tabular}{ll | ll | ll}
\hline 
& & \multicolumn{2}{c|}{{\it NB816}}  &  \multicolumn{2}{c}{{\it NB921}}   \\ 
Parameter  & Prior & Best-fit & Posterior PDF & Best-fit & Posterior PDF \\
\hline 
$\log{M_{\rm c}} / (h^{-1}M_\odot)$     &  None               & $11.75$ & $12.04^{+0.57}_{-0.32}$     & $11.93$ & $11.91^{+ 0.19} _{- 0.18}$ \\  
$\log{M_{\rm min}} / (h^{-1}M_\odot)$ & None                & $12.46$ & $12.61^{+0.32}_{-0.43}$    & $12.47$ & $12.57^{+0.29}_{-0.35}$ \\ 
$\sigma_{\log{M}}$                              & $[0,1]$             & $0.06$ & $0.40^{+0.28}_{-0.31}$ &        $0.13$ & $0.17^{+ 0.15}_{- 0.10}$  \\
$\alpha$                                               & $1.00\pm 0.20$  & $1.06$ & $1.03^{+0.12}_{-0.13}$         & $1.23$ & $1.12^{+ 0.13}_{- 0.12}$ \\
$F_c^A$                                               & $[0, 0.5]$         & $0.13$ & $0.26^{+0.16}_{-0.12}$         & 0.14 & $0.26^{+ 0.15}_{- 0.13}$  \\
$F_c^B$                                               & $[0,1]$         & $0.95$ & $0.37^{+0.39}_{-0.28}$        & 0.90 & $0.53^{+ 0.32}_{- 0.34}$  \\
$F_s$                                                  & $[0, 1]$              & $0.98$ & $0.54^{+0.31}_{-0.34}$         & 0.73 & $0.55^{+ 0.31}_{- 0.33}$  \\
$f_{\rm fake}$                                      & $0.140\pm 0.060$ & $0.172$ & $0.140^{+0.048}_{-0.053}$ & 0.128 & $0.104^{+ 0.043}_{- 0.041}$  \\
$\chi^2/\nu$  \ \ ($\nu=7$) & & 2.17& & 0.856& \\
\hline 
Inferred quantity & Measurement & Best-fit & Posterior PDF & Best-fit & Posterior PDF \\
\hline
$\log{n_g / (h^{-1}{\rm Mpc})^{-3}}$ &$-2.259\pm 0.068$ ({\it NB816}) & $-2.313 $ & $-2.316^{+0.071}_{-0.073}$&$-2.401$ & $-2.381^{+0.074}_{-0.073}$  \\
  & $-2.344\pm 0.070$ ({\it NB921}) & &  &      &  \\
$f_{\rm sat}$ & \multicolumn{1}{c |}{$\cdots$}                                             & $0.308$ & $0.159^{+0.120}_{-0.047}$ & $0.242$ & $0.159^{ + 0.109}_{ - 0.049}$ \\
$b_{\rm eff}$ & \multicolumn{1}{c |}{$\cdots$}                                             & $1.797$ & $1.700^{+0.084}_{-0.111}$ & $2.029$ & $1.981^{+0.072}_{-0.068}$ \\
$\log{M_{\rm eff}} / (h^{-1}M_\odot) $ & \multicolumn{1}{c |}{$\cdots$}       & $12.831$ & $12.703^{+0.093}_{-0.071}$ & $12.699$ & $12.609^{+0.085}_{-0.051}$  \\
\hline
\end{tabular}
\end{center}
\label{tab:hod_geach}
\begin{tabnote}
$*$ In the ``Prior'' column the ranges specified in brackets are for
  uniform priors while the others we quote the mean and standard
  deviation of the Gaussian priors. The column of ``Best-fit'' shows
  the parameter set which gives the minimum value of $\chi^2$. In the
  column of ``Posterior PDF'', the central value is a median and the
  error means $16-84$ percentiles after other parameters are
  marginalized over. 
  The measured number density $\log{n_g}$ includes non-[OII] emitters, 
  and thus its best-fitting values differ from the measure ones by $\log{(1-f_{\rm fake})} \sim -0.06$.
\end{tabnote}
\end{table*}

\begin{figure}[bt]
\begin{center}
\vspace{-.4cm}
\FigureFile(80mm,80mm){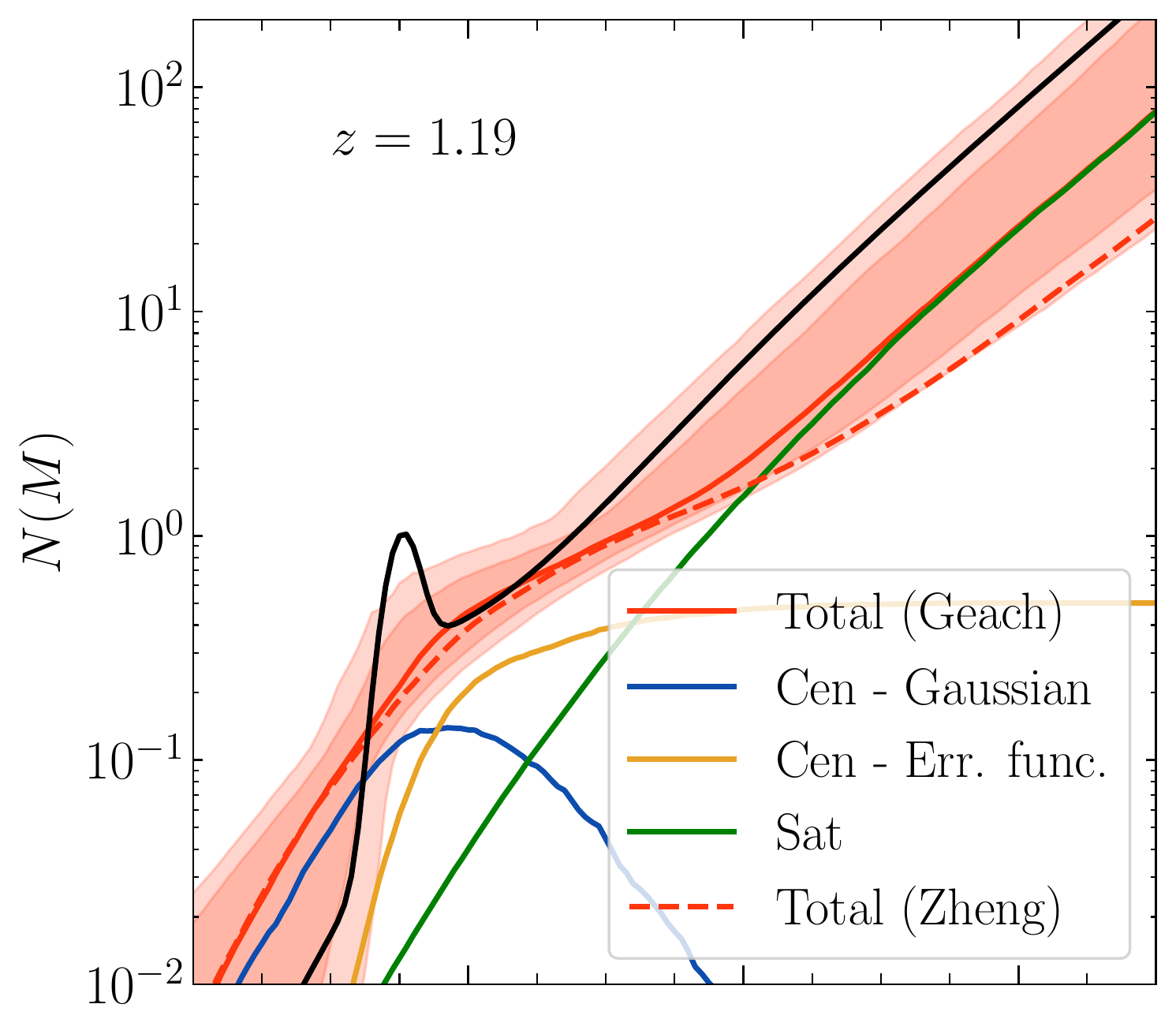} 
\\ \vspace{-.4cm}
\FigureFile(80mm,80mm){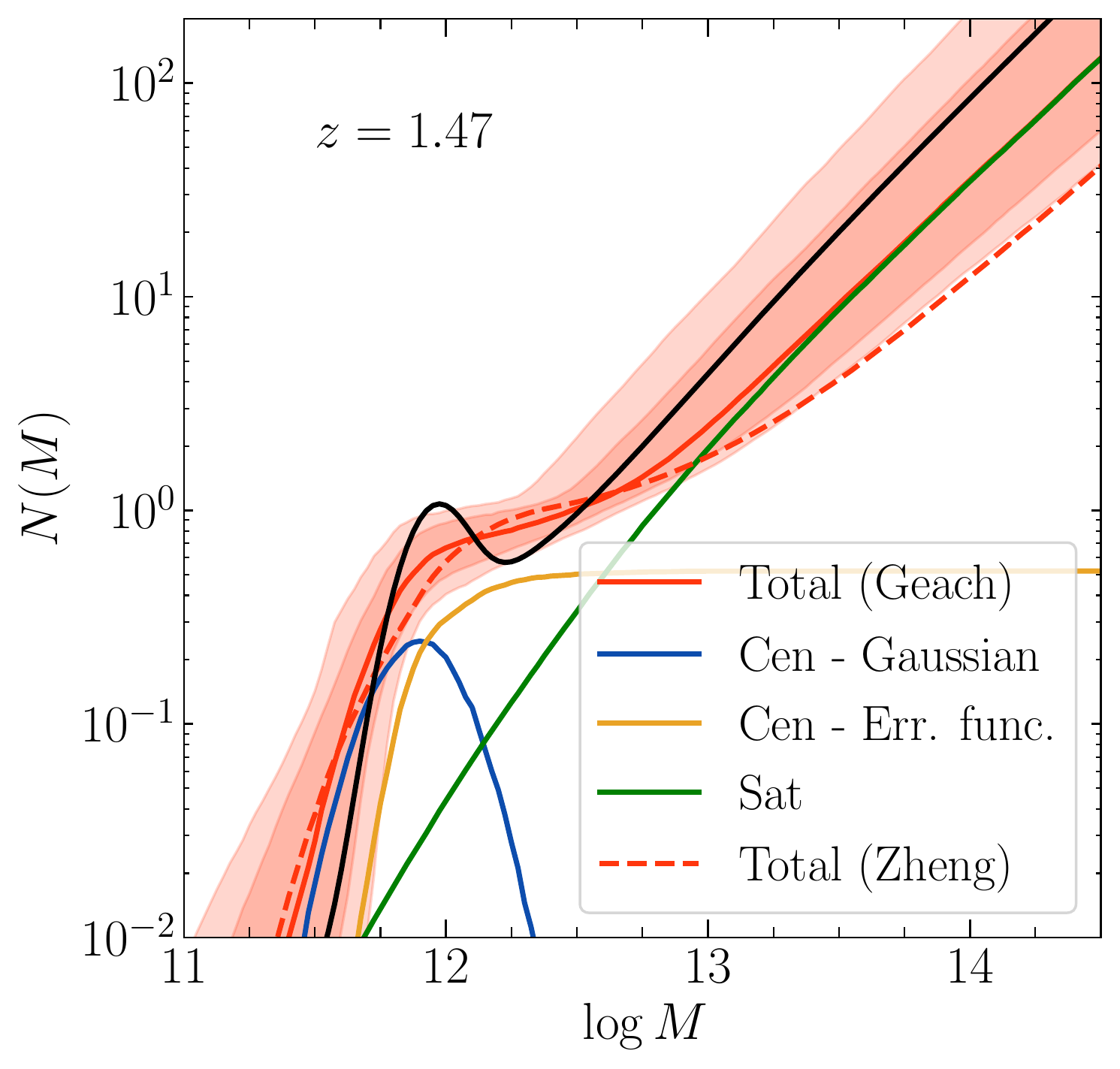}
\end{center}
\caption{HOD with the best-fitting parameters of Geach model at
  $z=1.19$ (top) and $z=1.47$ (bottom).  The blue and yellow lines
  show the average numbers of centrals from the Gaussian and error
  function terms (see equation (\ref{eq:hod_geach_cen})), while the
  green line shows that of satellites. The red solid line is the sum
  of the blue, yellow and green lines, the average number of all the
  [OII] emitters. The dark and light red shaded regions indicate the
  $68\%$ and $95\%$ confidence intervals.  The red dashed curve is the
  same as the red solid curve but for the best-fitting HOD of
  \citet{Zheng:2005}.  For comparison, the black curve shows the HOD
  of the Geach model which gives the minimum $\chi^2$ value.  }
\label{fig:result_hodfig_contreras}
\end{figure}

\begin{figure}[bt]
\begin{center}
\vspace{-.4cm}
\FigureFile(80mm,80mm){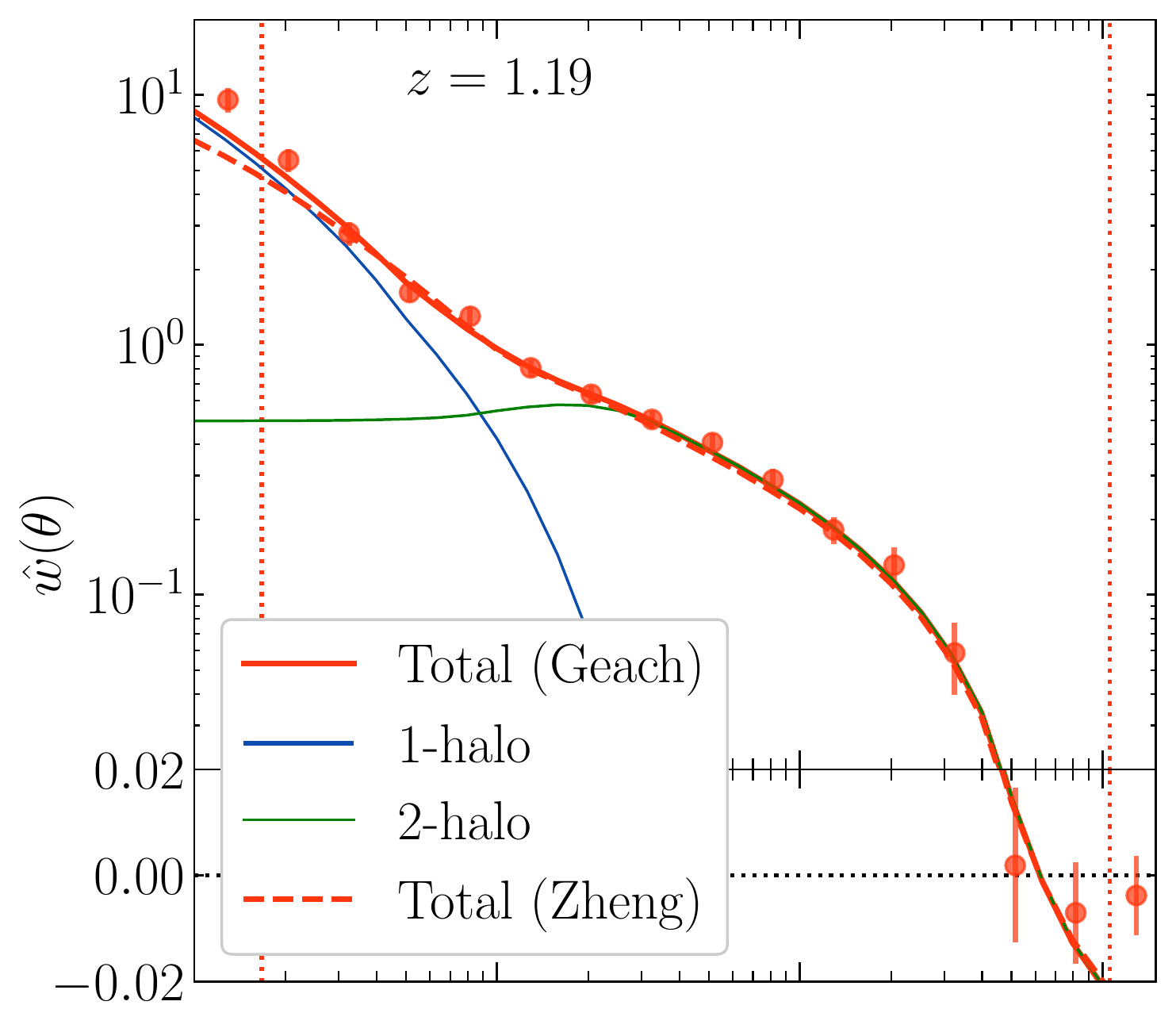}
\\ \vspace{-.4cm}
\FigureFile(80mm,80mm){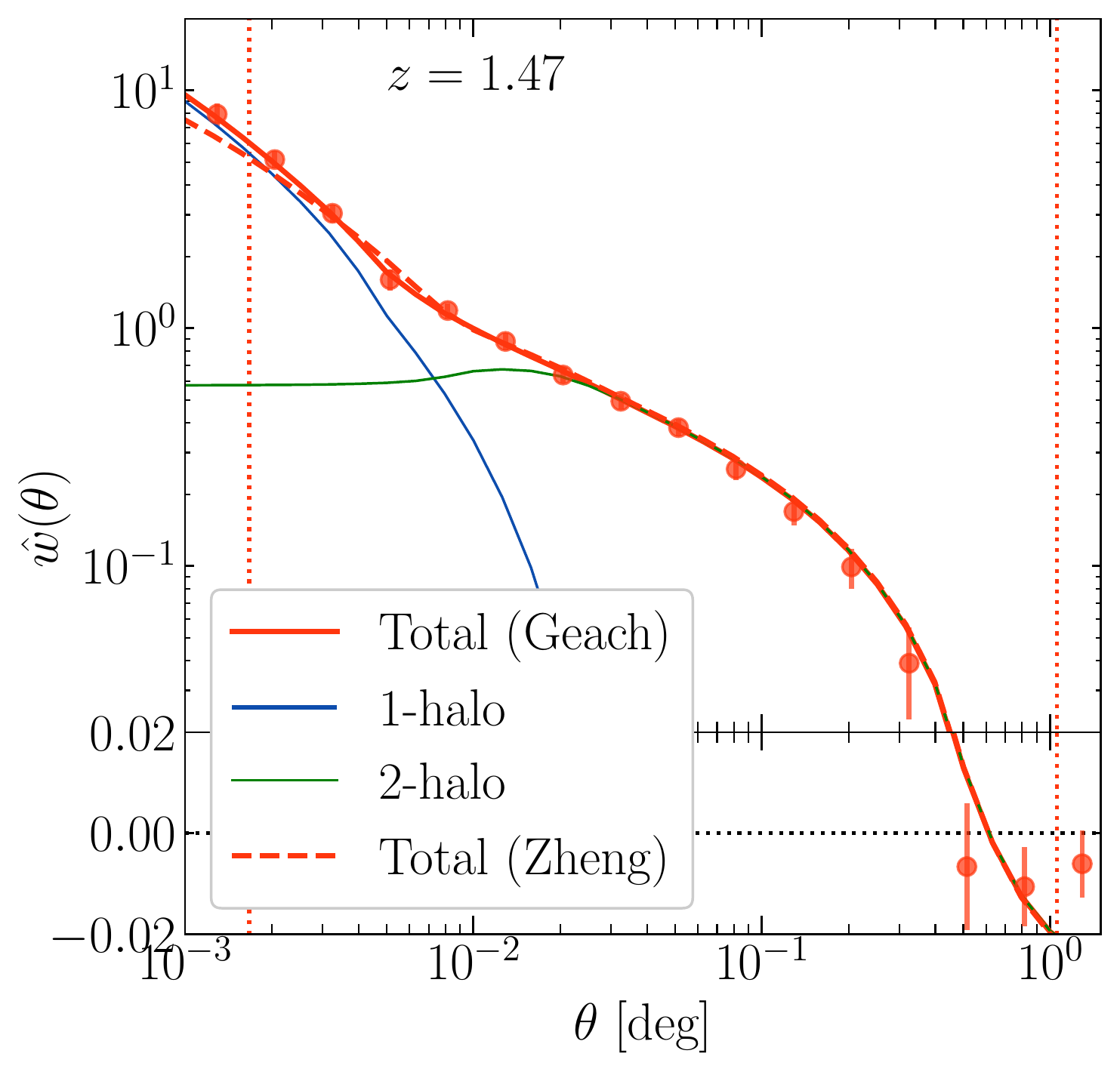}
\end{center}
\caption{The angular correlation functions of [OII] emitters at $z=1.19$ (top) and
  $z=1.47$ (bottom).  Note that the vertical axis mixes logarithmic
  and linear scalings.  The red points are the measurement, $\hat{w}$,
  the same as those in figure \ref{fig:w_all}.  The blue and green
  solid curves are the one- and two-halo terms with the best-fitting
  Geach HOD model, respectively, and the red solid curve is their
  sum. For comparison, we also show the best-fitting Zheng HOD model
  as the red dashed curve.  The data enclosed by the two red vertical
  lines are used for this HOD analysis.  }
\label{fig:result_xi_contreras}
\end{figure}

\subsection{Halo occupation distribution model for [OII] emitters} \label{sec:model}

In this paper we adopt the HOD model developed by \citet{Geach:2012}
to describe the population of ELGs (see also
\cite{Kim:2011,Contreras:2013b}). In this model, the central HOD is
described by two components:
\begin{eqnarray}
\left\langle N_{\rm cen}(M)\right\rangle &=& F_c^B(1-F_c^A) \exp{\left[ -\frac{\log{(M/M_{\rm c})}^2}{2(\sigma_{\log{M}})^2} \right]} \nn \\
&& + F_c^A \left [1+{\rm erf} \left(\frac{\log{(M/M_{\rm c})}}{\sigma_{\log{M}}}\right) \right], \label{eq:hod_geach_cen}
\eey
where $F_c^{A,B}$ are normalization factors. The first component
describes the Gaussian distribution of centrals around halos of
average mass $M_{\rm c}$ with the dispersion $\sigma_{\log{M}}$, and
the second component describes the standard mass-limited step function
form proposed by \citet{Zheng:2005} (see section \ref{sec:hod_zheng}
below).  The satellite HOD is given by
\be
\left\langle N_{\rm sat}(M) \right\rangle= F_s \left [1+{\rm erf} \left(\frac{\log{(M/M_{\rm min})}}{\delta_{\log{M}}}\right) \right] \left( \frac{M}{M_{\rm min}}\right)^\alpha, \label{eq:hod_geach_sat}
\ee
where $F_s$ is the mean number of satellites per halo at the
transition mass $M_{\rm min}$ which corresponds to the characteristic
mass above which halos can contain satellites, $\delta_{\log{M}}$ is
the width of the transition from zero satellites per halo to the power
law, and $\alpha$ is the slope of the power law which gives the mean
number of satellites for $M>M_{\rm min}$.  We refer the reader to
\citet{Geach:2012} and \citet{Contreras:2013b} for a more detailed
explanation of this model.

We fix the least-important parameter, $\delta_{\log{M}}$, to
$\delta_{\log{M}}=1$.  Thus the number of free parameters in the Geach HOD
model together with the fake line fraction is eight, $\Theta = (M_{\rm
  c}, M_{\rm min}, \sigma_{\log{M}}, \alpha, F_c^A, F_c^B, F_s, f_{\rm
  fake})$.  The parameter $\alpha$ is known to have a value around
unity.  We thus apply a Gaussian prior on $\alpha$ as $\alpha = 1.00
\pm 0.20$.  As with the analysis in section \ref{sec:result}, we
further impose a Gaussian prior on the fake line fraction parameter as
$f_{\rm fake}= 0.140 \pm 0.060$.  Uniform priors are applied for the
other 6 parameters.  Table \ref{tab:hod_geach} summarizes the priors
on all the eight parameters.  We use the Python package
\texttt{HALOMOD} \citep{Murray:2013} to calculate the HOD model
prediction for the angular correlation function.

\subsection{HOD parameter constraints} \label{sec:constraint}
We present constraints on the HOD model using the measured angular
correlation function of the [OII] emitters. We compute the $\chi^2$
statistic for the HOD model constructed from the observed angular
correlation function, $\chi_w^2$, and number density, $\chi_n^2$, as
\be
\chi^2 (\Theta) = \chi_w^2 (\Theta) +   \chi_n^2 (\Theta),
\ee
where $\chi_w^2$ is given by equation (\ref{eq:chi2_w}) and $\chi_n^2$
is given by
\be
 \chi_n^2 (\Theta) = \frac{\left[\log{n_g^{\rm obs}}-\log{n_g^{\rm th}(\Theta)}+\log{(1-f_{\rm fake})}\right]^2}{\sigma^2_{\log{n_g}}},
\ee
where $\Theta = (\log{M_{\rm c}}, \log{M_{\rm min}}, \alpha,
\sigma_{\log{M}}, F_s, F_c^A,F_c^B,f_{\rm fake})$.  Since the observed
number of [OII] emitters includes non-[OII] contaminants characterized
by the fraction $f_{\rm fake}$, we need to take into account a factor
of $(1-f_{\rm fake})$ difference between the observed number density
and its theoretical prediction computed by equation (\ref{eq:ng_hod}).
As shown in table \ref{tab:four_fields}, the observed number density
is $n_g^{\rm obs} = 5.50 \times 10^{-3} (h/{\rm Mpc})^3$ at $z=1.19$
and $n_g^{\rm obs} = 4.08 \times 10^{-3} (h/{\rm Mpc})^3$ at $z=1.47$.
We set the uncertainty of $\log{n_g^{\rm obs}}$ as $\sigma_{\log{n_g}}
= 0.03 |\log{n_g^{\rm obs}}|$ following the measurement uncertainty of
the luminosity function \citep{Hayashi:2020}.  Due to the flexibility
of the model, the HOD enables us to fit an observed correlation
function over a broader range than the power-law and nonlinear dark
matter models considered in the previous section.  We thus use the
data at $0.0015<\theta<1~[{\rm deg}]$ and the number of angular
separation bins is $N_{\rm bin} = 14$.  Together with the number
density, the degree of freedom is $\nu = N_{\rm bin} +1 - 8 = 7$.

The resulting constraints on the HOD parameters of the Geach model are
shown in figure \ref{fig:result_hod_contreras}.  This and the following
two figures are the main results of this paper.  The eight dimensional
posterior distributions are visualized in two-dimensional contours
with the other six parameters marginalized over.  The top row of each
column shows the one-dimensional posterior.  The constraints are
summarized in table \ref{tab:hod_geach}.  The normalizations
parameters, $F_c^{A,B}$ and $F_s$, are poorly constrained, which are
expected and consistent with \citet{Geach:2012} and \citet{Hong:2019}.
Overall, constraints from the $z=1.47$ sample are tighter than those
from the $z=1.19$ one.  Particularly, the marginalized distribution
for $\sigma_{\log{M}}$ for the $z=1.19$ sample is much wider and even
becomes bimodal.  Interestingly, this bimodality was seen by
\citet{Hong:2019} who analyzed the clustering of Ly$\alpha$ emitters
using the Geach HOD model.  Although there are many free parameters
with few priors, we obtain meaningful constraints due to 
our large sample. Particularly
strong constrains are obtained for the two mass parameters, $M_{\rm
  c}$ and $M_{\rm min}$ as seen in table \ref{tab:hod_geach}, because
$M_c$ can be determined by the observed number density, $n_g$, and
$M_{\rm min}$ primarily degenerates with $M_{\rm c}$.  We do not see a
strong evolution of HOD from $z=1.47$ to $1.19$, which is consistent
with our ongoing work based on cosmological hydrodynamical simulations
\citep{Osato:2021}.

The posterior distribution of the HOD is shown in figure
\ref{fig:result_hodfig_contreras}.  The dark and light shaded regions
show the 68 and 95\% confidence intervals for the total HOD,
respectively, and red solid curve is the median. The black solid curve
is the HOD with a set of parameters which give the minimum $\chi^2$
value in the eight dimensional parameter space. The HOD with the
minimum $\chi^2$ looks somewhat different from the median of the HOD,
particularly the central HOD. This is a similar trend with the finding
of \citet{Hong:2019}. They found that two independent algorithms give
very different best-fitting HOD parameters (See Model\#1 and Model\#2
of their figure 7). This small but non-negligible discrepancy largely
comes from the fact that the normalization parameters, especially
$F_c^B$, and the scatter of the mass, $\sigma_{\log{M}}$, are poorly
constrained.  Even though the HODs are different, the predicted
angular correlation functions with these HOD parameter sets become
very similar.

The best-fitting Geach HOD model prediction, $(1-f_{\rm
  fake})^2(w(\theta;\Theta)-w_\Omega)$, is shown as the red solid
curve in figure \ref{fig:result_xi_contreras}. The contributions from
the 1-halo and 2-halo components are shown as the blue and green
curves, respectively.  Obviously, agreement of the measured
correlation function with the HOD model prediction is much more
remarkable than that with the power-law or linearly biased dark matter
model, compared to figure \ref{fig:w_all}.

\subsection{Derived physical parameters for host halos}

Using the constrained HOD parameters, one can infer the parameters
which characterize the properties of [OII] emitters and their host
dark matter halos, such as the galaxy number density, effective halo
bias and mass, and satellite galaxy fraction, through equations
(\ref{eq:ng_hod}), (\ref{eq:b_hod}), (\ref{eq:m_hod}), and
(\ref{eq:fsat_hod}), respectively.  The posterior distribution for
these parameters is shown in figure \ref{fig:result_params_b_m_n_f}
and summarized in table \ref{tab:hod_geach}.
%
%
The number density is constrained following the imposed prior. Note
again that the constrained number density is different from the
observed number density by a factor of $(1-f_{\rm fake})$, $n_g^{\rm
  th} = (1-f_{\rm fake})n_g^{\rm obs}$.
%
%
The effective bias parameter is determined at $z=1.19$ and $1.47$ as
$b_{\rm eff}=1.701^{+0.083}_{-0.110}$ and $b_{\rm
  eff}=1.981^{+0.072}_{-0.068}$, respectively.  They are fully
consistent with the linear bias parameter from much simpler analysis
in section \ref{sec:nl_matter}, $b=1.61^{+0.13}_{-0.11}$ and
$b=2.09^{+0.17}_{-0.15}$.

\begin{figure}
\begin{center}
\vspace{-.2cm}
\hspace{-0.2cm}
\FigureFile(20.0mm,45mm){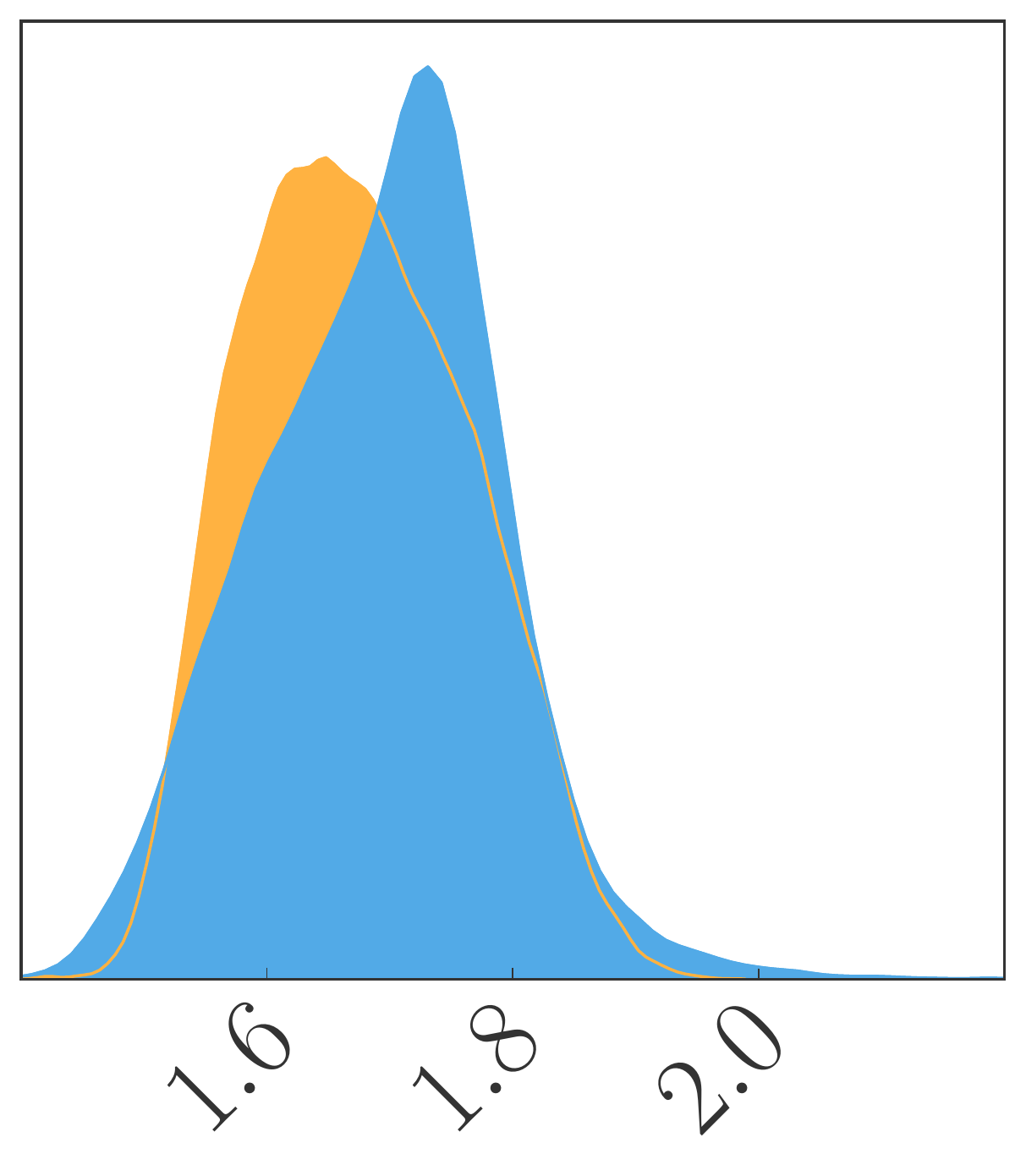}
\hspace{-0.05cm}
\FigureFile(20.0mm,45mm){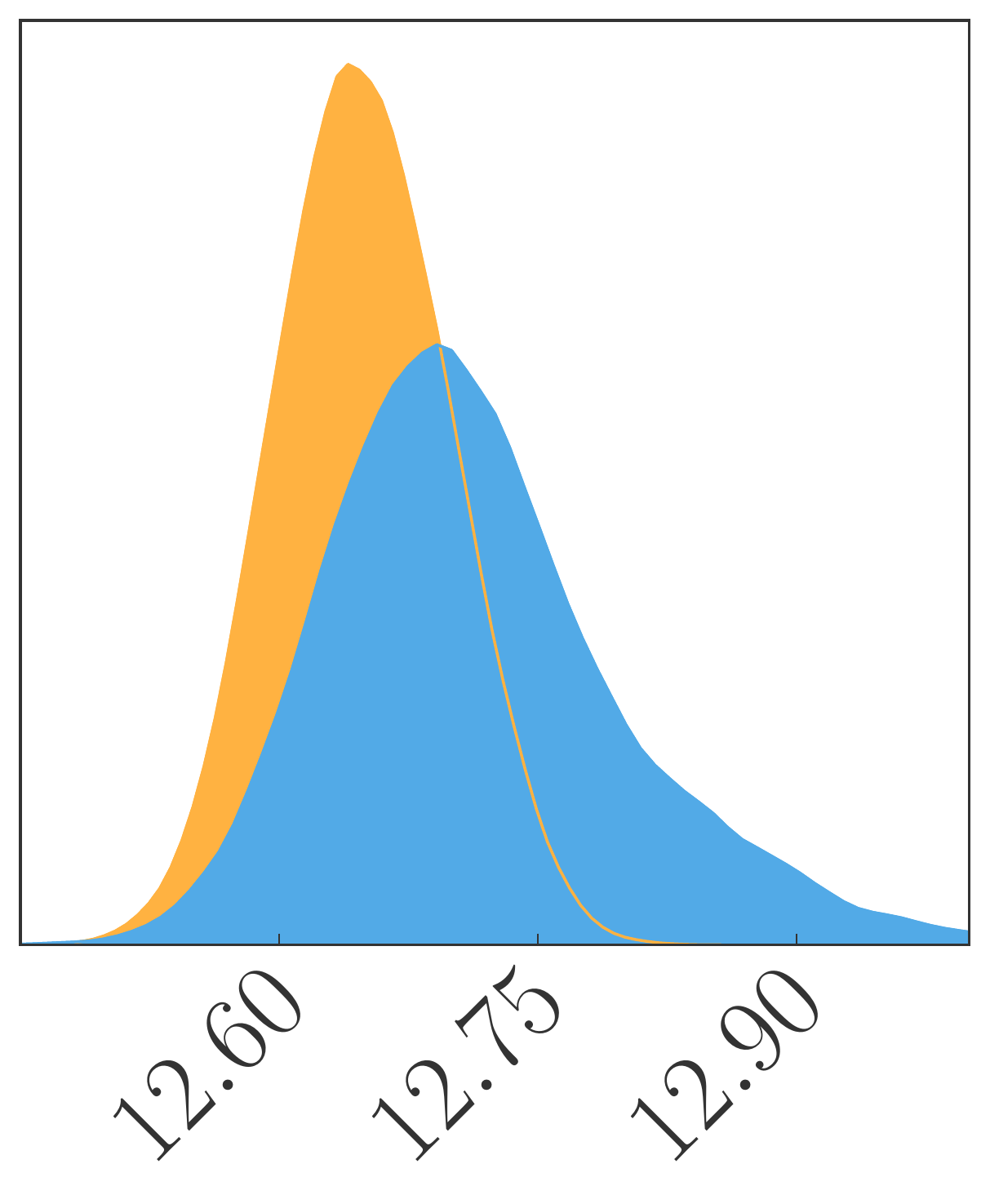}
\hspace{-0.05cm}
\FigureFile(20.0mm,45mm){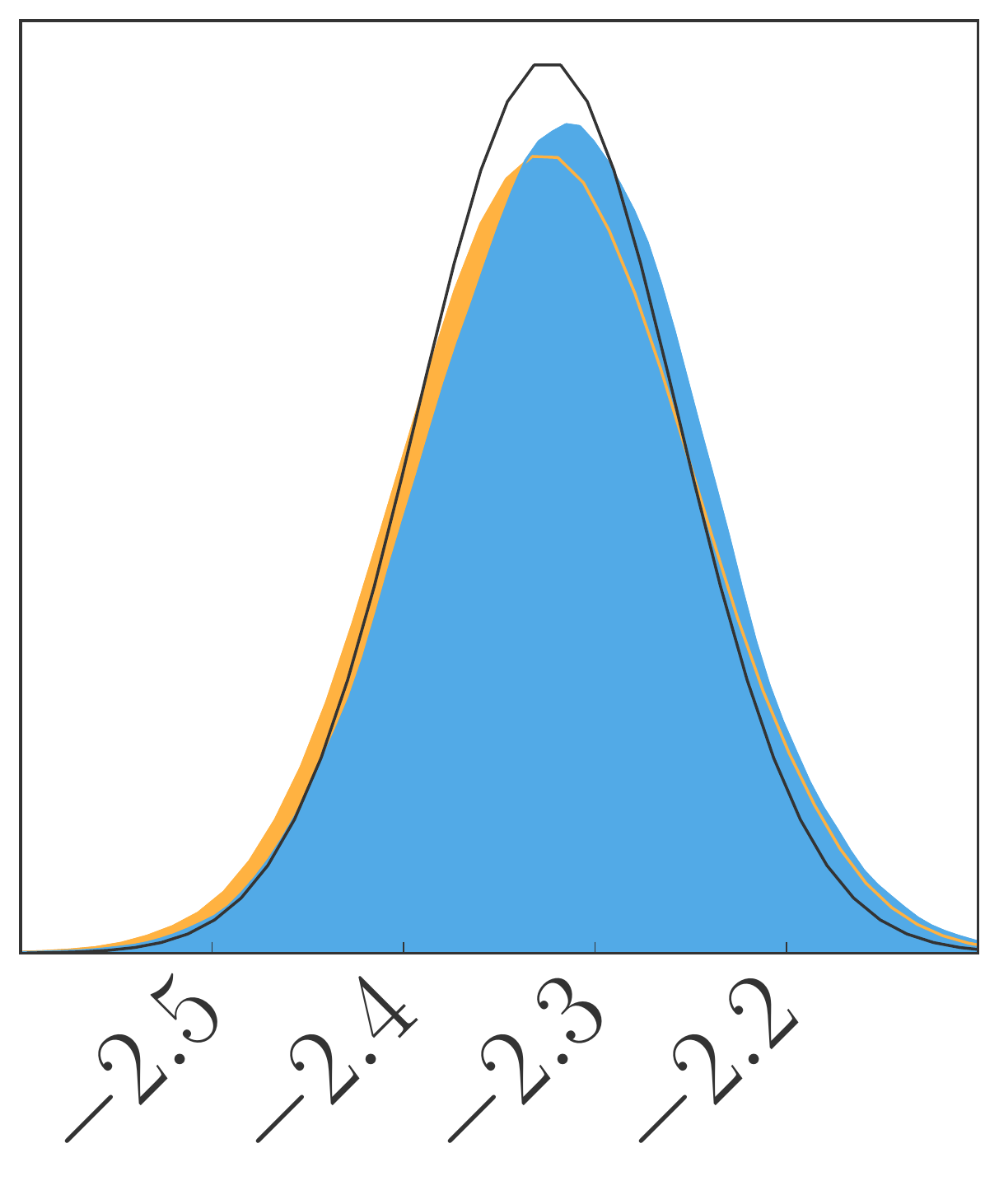}
\hspace{-0.18cm}
\vspace{-.35cm}
\FigureFile(21.2mm,45mm){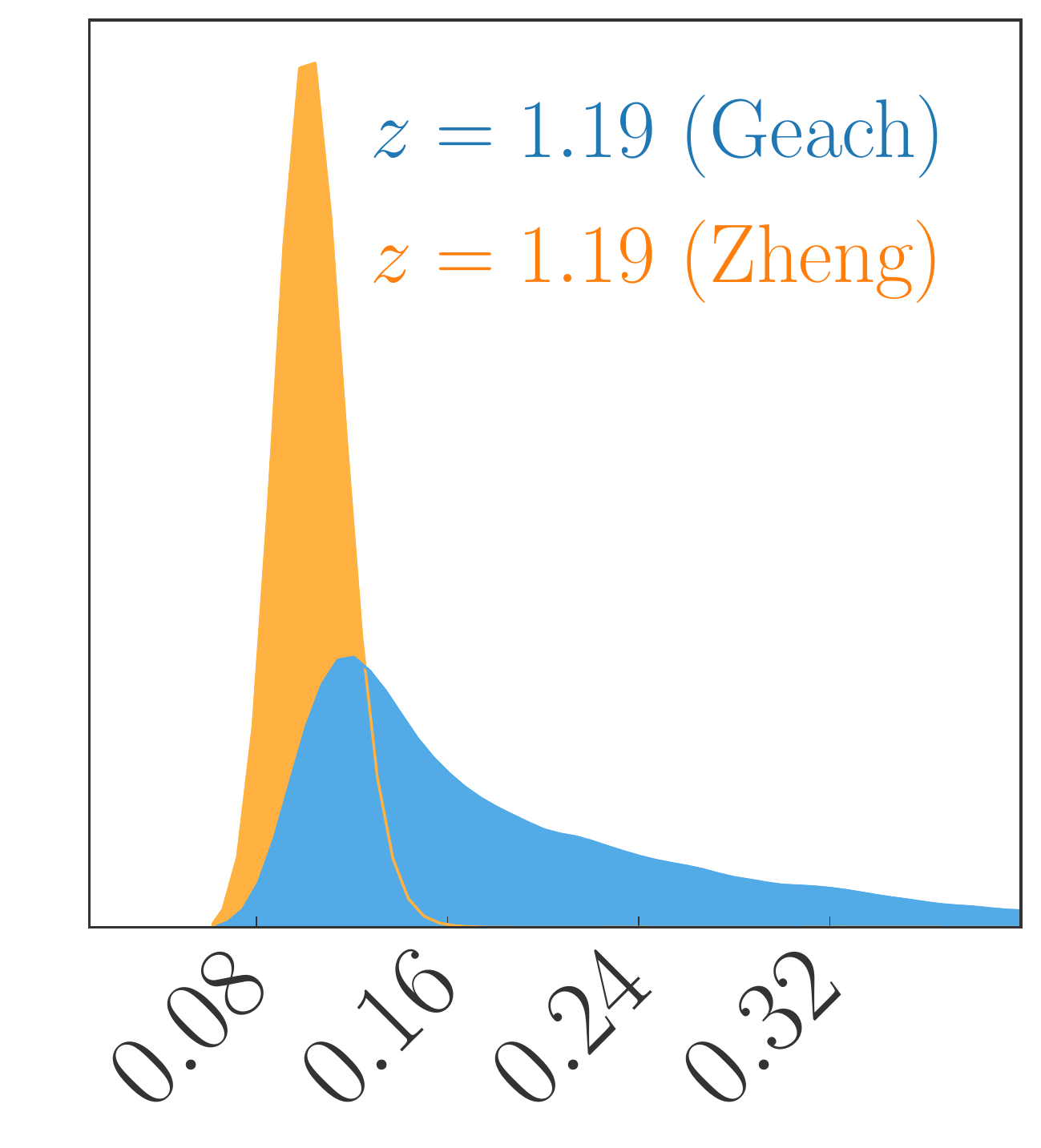}
\\
\hspace{-0.2cm}
\FigureFile(20.0mm,45mm){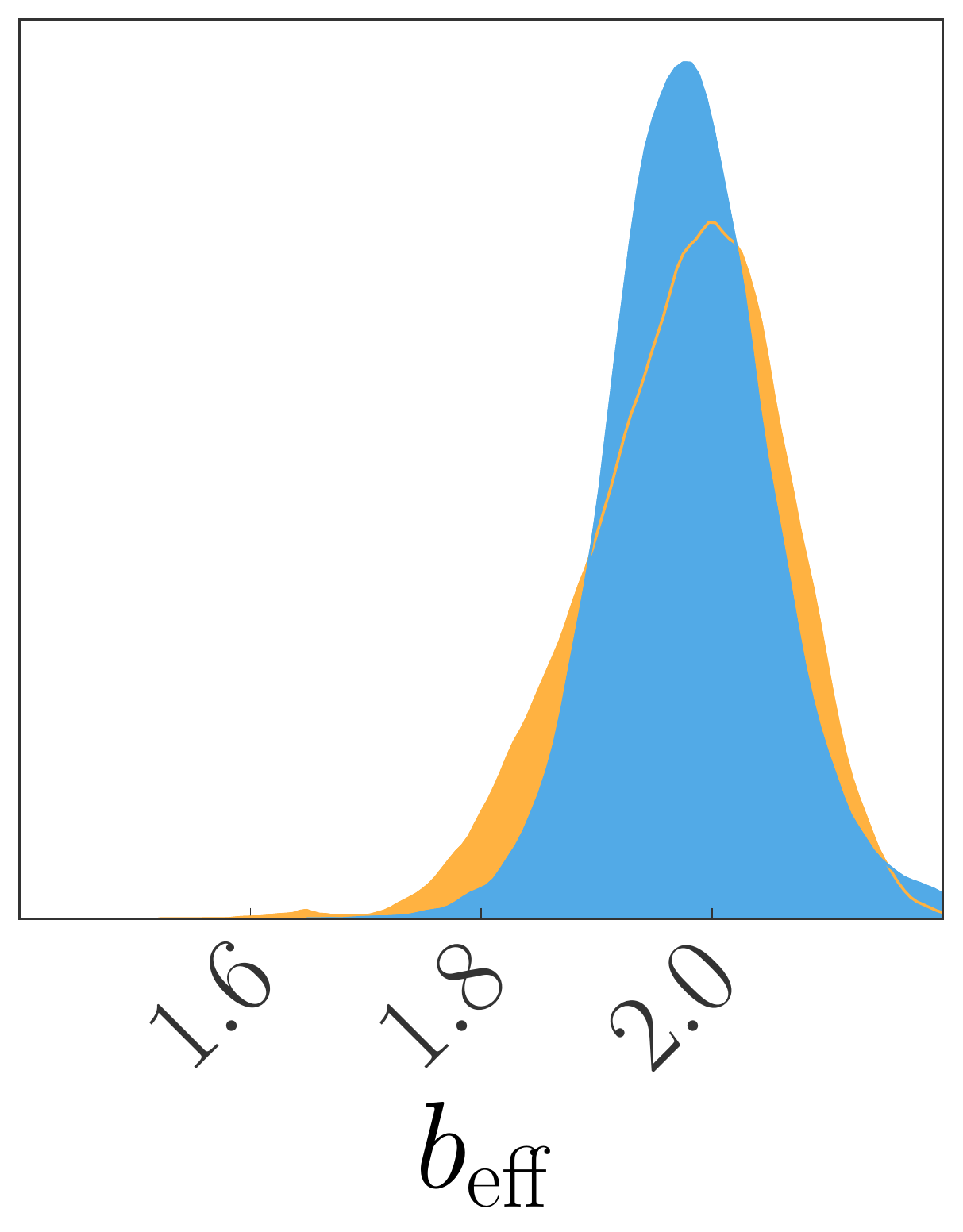}
\hspace{-0.05cm}
\FigureFile(20.0mm,45mm){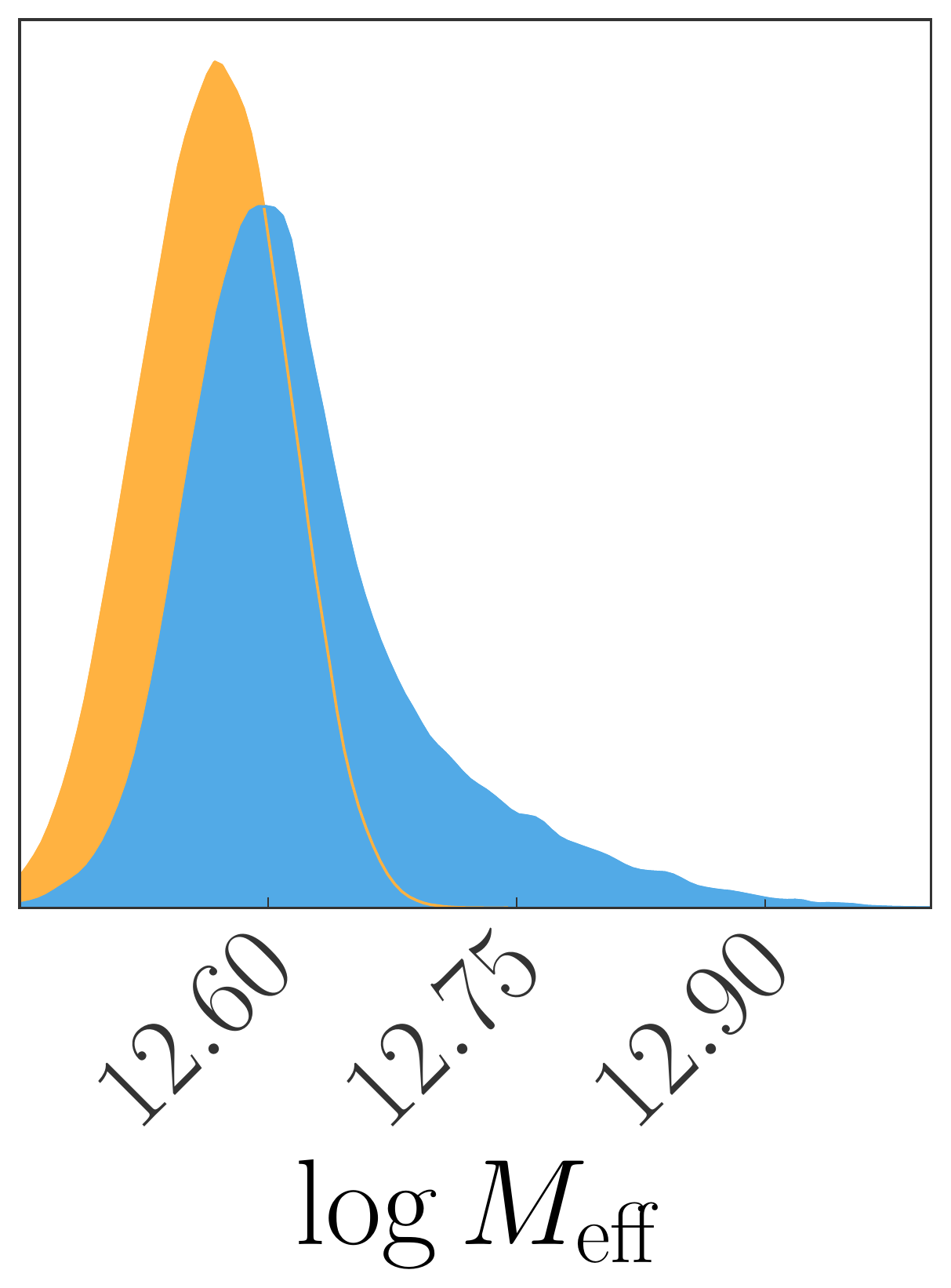}
\hspace{-0.05cm}
\FigureFile(20.mm,45mm){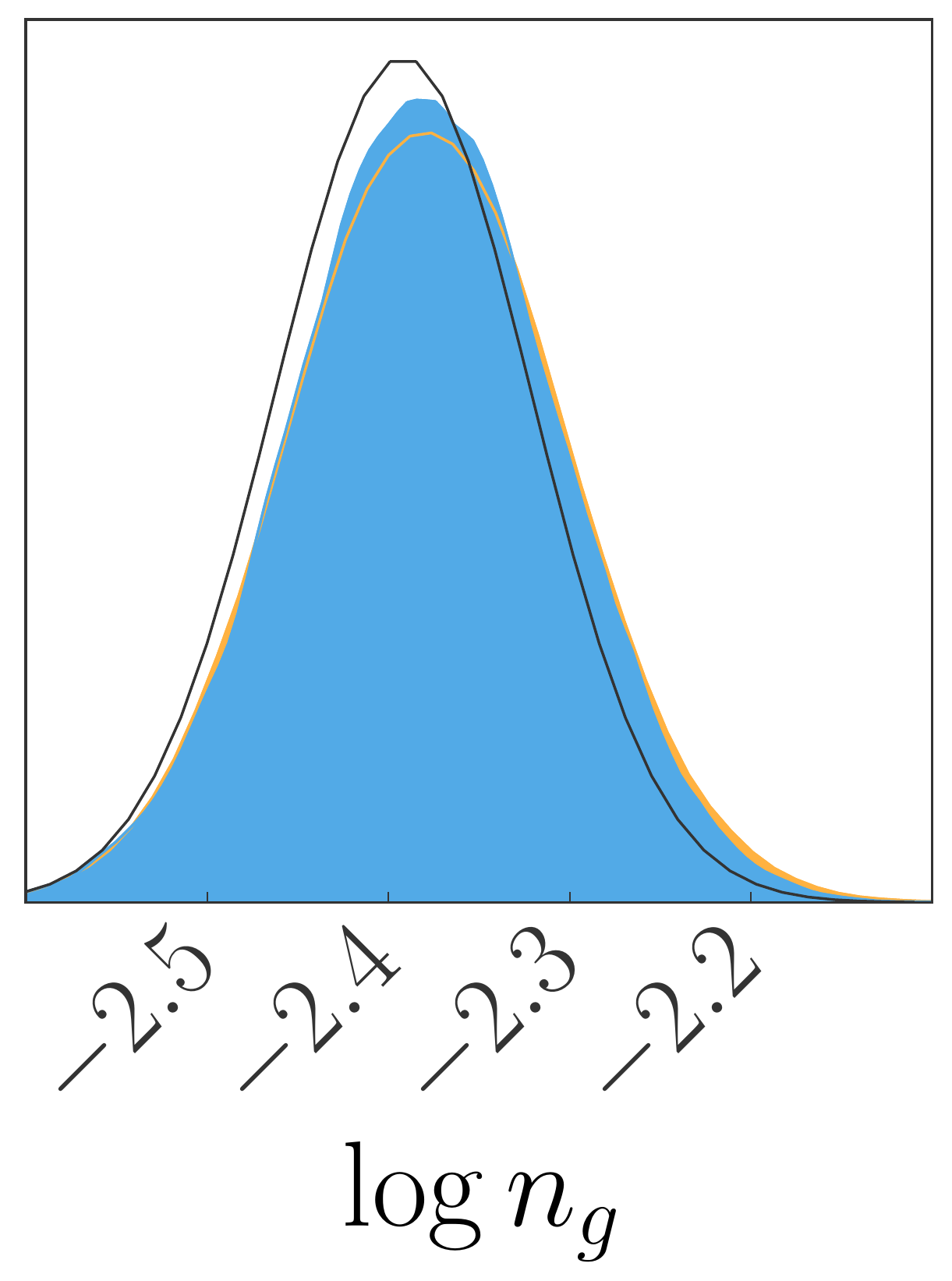}
\hspace{-0.18cm}
\FigureFile(21.2mm,45mm){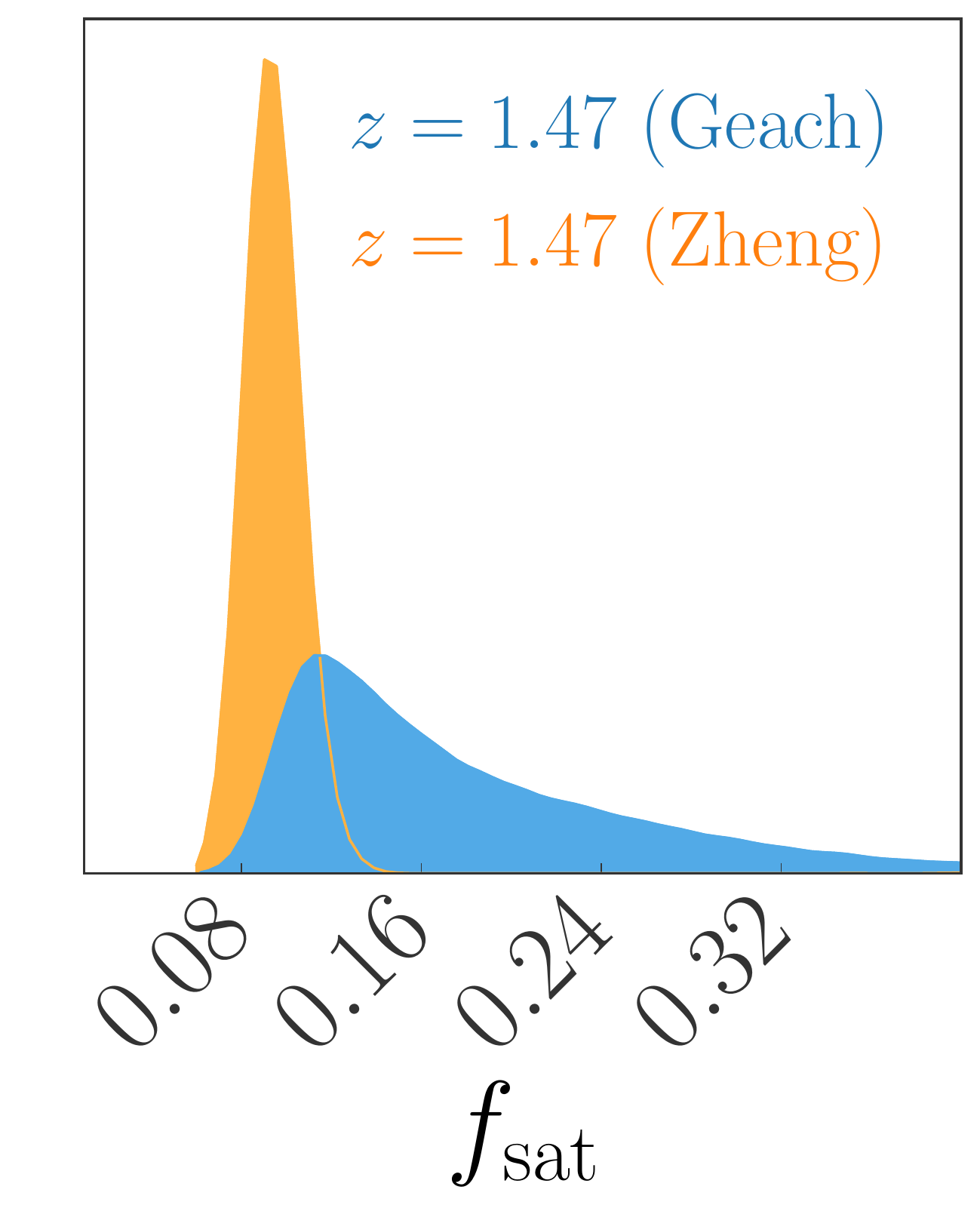}
\vspace{-0.3cm}
\end{center}
\caption{Posterior distribution for the parameters derived from the
  best-fitting parameters for the Geach (blue) and Zheng (orange) HOD
  models.  From the left to right, the posteriors for the effective
  bias, effective halo mass, the galaxy number density and the
  satellite fraction are shown. The upper and lower panels show
  results for $z=1.19$ and $z=1.47$, respectively.  
  The number density presented here is different from the observed number density 
  shown in table \ref{tab:four_fields} by a factor of $(1-f_{\rm fake})$, $(1-f_{\rm fake})n_g^{\rm obs}$ 
  (see table \ref{tab:hod_geach} for the constrained values for $f_{\rm fake}$). 
  Gaussian priors assumed for $\log{n_g}$ are depicted by the black solid curves, where the best-fitting value of $f_{\rm fake}$ is adopted to multiply by $(1-f_{\rm fake})$.
  The posterior of $\log{n_g}$ for $z = 1.47$ (orange) is almost entirely behind the one for $z = 1.19$.
  }
\label{fig:result_params_b_m_n_f}
\end{figure}

%
%
The effective masses of halos hosting [OII] emitters are derived to be
$\log{M_{\rm eff}/(h^{-1}M_\odot)}=12.703^{+0.091}_{-0.069}$ and
$12.609^{+0.085}_{-0.051}$ at $z=1.19$ and $1.47$, respectively.
Figure \ref{fig:mass_accretion} shows the constraints on $M_{\rm eff}$
as a function of redshift obtained from our HSC [OII] emitters
together with the previous studies at $0<z<4$ which were already
presented in \citet{Kashino:2017a}.  The constraints include the
studies of photo-$z$ galaxies from the CFHT Legacy Survey at $z\sim 0.3$,
$0.5$ and $0.7$ \citep{Coupon:2012}, VIMOS-VLT Deep Survey
(VVDS) at $z\sim 0.55$ and $1$ \citep{Abbas:2010}, NEWFIRM Medium Band
Survey at $z\sim 1.1$ and $1.5$ \citep{Wake:2011}, H$\alpha$ emitters
from the FMOS-COSMOS Survey at $z\sim 1.6$ \citep{Kashino:2017a} and
from the HiZELS survey at $z\sim 2.2$ \citep{Geach:2012}, and VUDS at
$z\sim 2.5$ and $3.5$ \citep{Durkalec:2015}.  These samples have
number densities and stellar masses similar to our [OII] emitter
samples.
Here we also plot the average mass assembly history of halos with
different present-day masses, as derived by \citet{Behroozi:2013a}
based on an $N$-body simulation.  According to this prediction, our
measurements of $M_{\rm eff}$ at two redshifts are well explained by
the mass-assembly history with $M(z=0)=1.5\times 10^{13}M_\odot / h$
as depicted by the blue solid curve.  The effective halo mass of [OII]
emitters at $z=1.47$ is slightly higher than the blue curve, which
reflects the fact that [OII] emitters of our $z=1.47$ sample are
intrinsically brighter than those of the lower-z sample (see section
\ref{sec:data_nb}).
Given the accurate constraints on HOD, one can in principle infer the
mass of the host halo for an individual emission line galaxy
\citep{Oguri:2015}. It is, however, beyond the scope of this paper and
will be investigated in future work.

%
%
For our [OII] emitter sample, the satellite galaxy fraction is
constrained as $f_{\rm sat}=0.158^{+0.114}_{-0.047}$ and
$0.159^{+0.109}_{-0.049}$ at $z=1.19$ and $1.47$, respectively.  The
preceding studies revealed that the satellite galaxy fraction of a
given population would depend on the redshift, number density and
stellar mass \citep{Coupon:2012,Guo:2014,Guo:2019}.  At least over the
redshift we studied here, the satellite fraction of [OII] emitters in
our sample does not significantly evolve with redshift.
The analysis of \citet{Favole:2016} constrained the satellite fraction
of ELGs with the host halo mass $M\sim 10^{12}h^{-1}M_\odot$ as
$f_{\rm sat}=0.225\pm 0.025$.
\citet{Avila:2020} constrained the value of $f_{\rm sat}$ for ELGs
from the eBOSS survey, and the best fitting value of their baseline
model is $f_{\rm sat} = 0.22$, consistent with our
measurements. However, their constraints are significantly
model-dependent and varying the baseline model changes the
best-fitting value between $0.18\leq f_{\rm sat}\leq 0.70$. 
Moreover, the selection of ELGs in the eBOSS survey is quite 
different from ours based on the NB. We thus
cannot make a quantitative comparison with their result.
\citet{Guo:2019} further studied the relation among the satellite
fraction, stellar mass and halo mass for ELGs from the eBOSS survey.
They showed that the host halo mass of ELGs with $f_{\rm sat} \sim
0.15$ is $\log{M/(h^{-1}M_\odot)}\sim 12.6$ is a strong function of
the stellar mass. It provides good agreement with our clustering
measurements of [OII] emitters from the HSC survey.

\begin{figure}[bt]
\begin{center}
\vspace{-.3cm}
\FigureFile(82mm,82mm){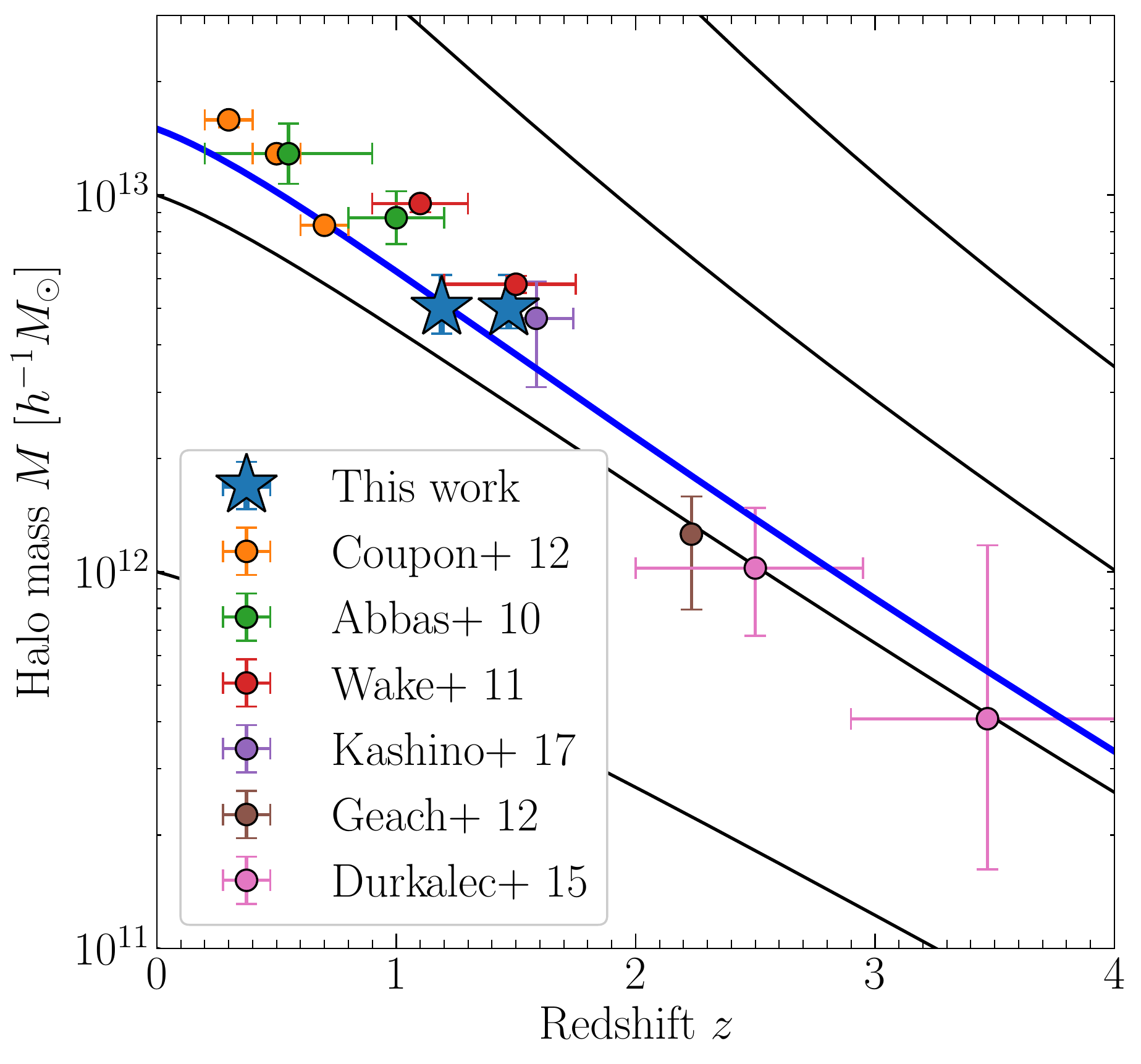}
\end{center}
\caption{Average mass of halos hosting ELGs as a function of redshift.
  The blue stars are the result from our analysis of [OII] emitters at
  $z=1.19$ and $z=1.47$.  The other points are the results of the
  previous HOD analysis by \citet{Abbas:2010}, \citet{Wake:2011},
  \citet{Geach:2012}, \citet{Coupon:2012}, \citet{Durkalec:2015}, and
  \citet{Kashino:2017a} (see the legend).  The solid curves show a
  prediction of mass accretion histories for halos with
  $M(z=0)=10^{15},10^{14},10^{13}$ and $10^{12}M_\odot$ from the top
  to bottom \citep{Behroozi:2013a}. The blue curve represents the one
  with $M(z=0)=1.5\times 10^{13}~[ h^{-1}M_\odot]$.  }
\label{fig:mass_accretion}
\end{figure}

\begin{figure*}
\begin{center}
\vspace{-.5cm}
\FigureFile(125mm,125mm){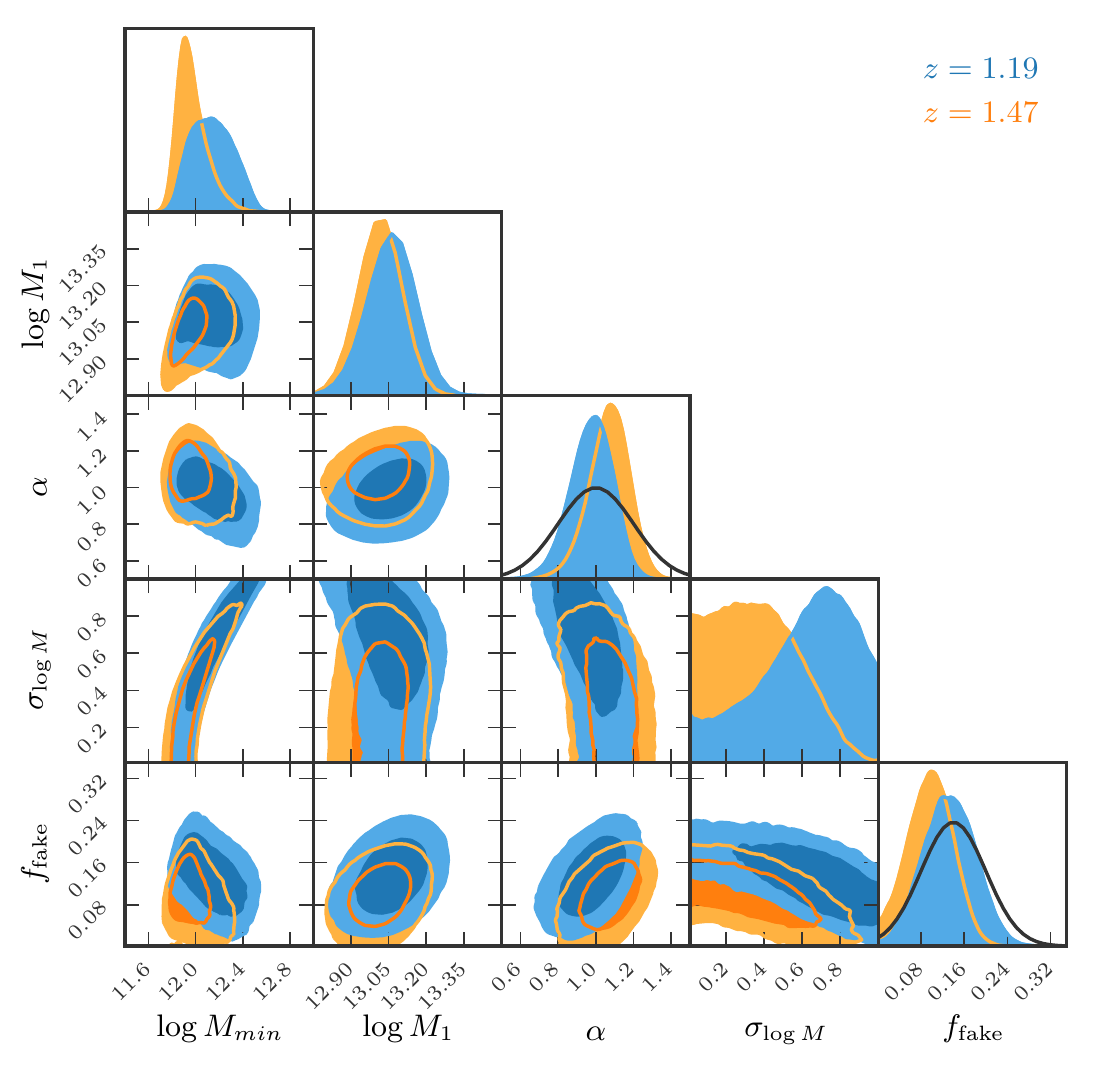}
\end{center}
\caption{Constraints on HOD parameters of Zheng's model, $(\log{M_{\rm
      c}}, \log{M_{\rm min}}, \alpha, \sigma_{\log{M}}) $ and $f_{\rm
    fake}$ for $z=1.19$ (blue) and $z=1.47$ (orange). Contours show
  the $68\%$ and $95\%$ confidence levels. The diagonal panels show the
  posterior probability distribution of each parameter. Gaussian
  priors are assumed for $\alpha$ and $f_{\rm fake}$, as depicted by
  the black curves in the panels of the 1-d posterior distributions.
}
\label{fig:result_hod_zheng}
\end{figure*}

\subsection{Comparison to a simpler HOD}\label{sec:hod_zheng}
While the Geach HOD model is general and flexible, as a trade-off, the
HOD parameters are constrained more poorly and more degenerated with
each other.  
Furthermore, it is not clear whether different halo occupation functions provide the same physical interpretation for a measured correlation function. 
A similar attempt has been made by studying one of the simplest HOD model
\citep{Zehavi:2005} in \citet{Hong:2019} (see their Appendix B).  Here
we investigate whether the predicted correlation function, halo
occupation function, and the host halo parameters of [OII] emitters
can be different between different HOD models.

To see this, we consider the commonly-adopted model of
\citet{Zheng:2005} (see also \cite{Zheng:2007}).  The model contains five parameters in total and
the central and satellite HODs are, respectively, given by
\begin{eqnarray}
  && N_{\rm cen}(M) =
  \frac{1}{2}\left[ 1+ {\rm erf}\left(\frac{\log{M / M_{\rm min}}}
  {\sigma_{\log{M}}}\right) \right],  \\ 
  && N_{\rm sat}(M) = 
   \frac{1}{2}\left[ 1+ {\rm erf}\left(\frac{\log{M / M_{\rm min}}}
  {\sigma_{\log{M}}}\right) \right]
  \left(\frac{M-M_0}{M_1}\right)^{\alpha},
\end{eqnarray}
where $M_{\rm min}$ is the characteristic mass to host a central
galaxy, $M_1$ is a mass for a halo with a central galaxy to host one
satellite, $M_0$ is the mass scale to truncate satellites,
$\sigma_{\log{M}}$ is the characteristic transition width, and
$\alpha$ is the slope of the power law for the satellite HOD, the same
as $\alpha$ in the Geach model.  The parameter $M_0$ is a
poorly-constrained parameter.  In order to make this analysis simpler,
we impose a relation between $M_0$ and $M_1$ following
\citet{Conroy:2006} (see also \cite{Kashino:2017a}),
\begin{equation}
\log{M_0} / (h^{-1}M_\odot) = 0.76 \log{M_1}/(h^{-1}M_\odot) + 2.3.
\end{equation}
Thus the number of free parameters in the Zheng HOD model together
with the fake line fraction is five, $\Theta = (M_{\rm min}, M_1,
\sigma_{\log{M}}, \alpha, f_{\rm fake})$.  As is the case of the
analysis with the Geach HOD model, we apply the Gaussian priors on
$\alpha$ and $f_{\rm fake}$ as $\alpha= 1.00\pm 0.20$ and $f_{\rm
  fake} = 0.140\pm 0.060$ , and flat priors on the other 3 parameters.
We use the data with the same angular separation range as the analysis
of the Geach HOD model (section \ref{sec:constraint}),
$0.0015 <\theta < 1$ [deg], the degree of freedom is
$\nu = N_{\rm bin} +1 - 5= 10$.

\begin{table*}[bt!]
\caption{Priors and constraints of the HOD parameters for Zheng model${}^*$}
\begin{center}
\begin{tabular}{ll | ll | ll}
\hline 
& & \multicolumn{2}{c|}{{\it NB816}}  &  \multicolumn{2}{c}{{\it NB921}}   \\ 
Parameter  & \multicolumn{1}{l |}{Prior} & Best-fit & Posterior PDF & Best-fit & Posterior PDF \\
\hline 
$\log{M_{\rm min}} / (h^{-1}M_\odot)$ & \multicolumn{1}{l |}{None}          & $12.16$ & $12.14^{+0.19}_{-0.18}$ &         $11.88$ & $11.95^{+0.15}_{-0.10}$ \\ 
$\log{M_1} / (h^{-1}M_\odot)$             & \multicolumn{1}{l |}{None}       & $13.070$ & $13.062^{+0.084}_{-0.090}$ & $13.000$ & $13.017^{+0.075}_{-0.078}$ \\ 
$\sigma_{\log{M}}$                              & \multicolumn{1}{l |}{$[0,1]$}            & $0.68$ & $0.65^{+0.21}_{-0.32}$ &             $0.18$ & $0.34^{+ 0.25}_{- 0.23}$  \\
$\alpha$                                               & \multicolumn{1}{l |}{$1.00\pm 0.20$}     & $0.99$ & $0.98^{+0.11}_{-0.12}$ &             $1.11$ & $1.07^{+ 0.10}_{- 0.11}$ \\
$f_{\rm fake}$                                      & \multicolumn{1}{l |}{$0.140\pm 0.060$} & $0.123$ & $0.133^{+0.048}_{-0.048}$  &    $0.115$ & $0.098^{+ 0.041}_{- 0.043}$  \\
$\chi^2/\nu$ \ \ ($\nu=10$)& & $1.86$& &$1.42$ & \\
\hline 
Inferred quantity & \multicolumn{1}{l |}{Measurement}  & Best-fit & Posterior PDF & Best-fit & Posterior PDF \\
\hline
$\log{n_g / (h^{-1}{\rm Mpc})^{-3}}$ & $-2.259\pm 0.068$ ({\it NB816})   & $-2.327$   & $-2.325^{+0.075}_{-0.073}$ &    $ -2.369$ & $-2.379^{+ 0.075}_{- 0.074}$ \\
  & $-2.344\pm 0.070$ ({\it NB921}) & &  &      &  \\
$f_{\rm sat}$ & \multicolumn{1}{c |}{$\cdots$}                                            & $0.100$ & $0.102^{+0.013}_{-0.012}$ &     $0.098$ & $0.094^{ + 0.012}_{ - 0.011}$ \\
$b_{\rm eff}$ & \multicolumn{1}{c |}{$\cdots$}                                            & $1.659$   & $1.668^{+0.102}_{-0.089}$ &           $2.034$ & $1.983^{+0.077}_{-0.095}$ \\
$\log{M_{\rm eff}} / (h^{-1}M_\odot) $ & \multicolumn{1}{c |}{$\cdots$}      & $12.645$ & $12.647^{+0.052}_{-0.050}$ & $12.591$ & $12.567^{+0.043}_{-0.046}$  \\
\hline
\end{tabular}
\end{center}
\label{tab:hod_zheng}
\begin{tabnote}
$*$ In the ``Prior'' column the ranges specified in brackets are for
  uniform priors while the others we quote the mean and standard
  deviation of the Gaussian priors. The column of ``Best-fit'' shows
  the parameter set which gives the minimum value of $\chi^2$. In the
  column of ``Posterior PDF'', the central value is a median and the
  error means $16-84$ percentiles after other parameters are
  marginalized over.
  The measured number density $\log{n_g}$ includes non-[OII] emitters, 
  and thus its best-fitting values differ from the measure ones by $\log{(1-f_{\rm fake})} \sim -0.06$.
\end{tabnote}
\end{table*}

The resulting constraints on the HOD parameters of Zheng's model are
shown in figure \ref{fig:result_hod_zheng} and summarized in table
\ref{tab:hod_zheng}.  Overall, the constraints on the Zheng HOD model are
tighter than those on the Geach model due to the fewer number of
parameters.  The posterior distribution of the HOD $N(M)$ is shown as
the red dashed curve in figure \ref{fig:result_hodfig_contreras}.
Interestingly, the two models predict very similar HODs, particularly
the central ones, for the observed angular correlation function of
[OII] emitters.  The discrepancy seen at $M>10^{13}M_\odot/h$ is
reasonable because such massive halos are rare as hosts of [OII]
emitters and since the varied range of $F_s$ in the Geach model is
$0\leq F_s \leq 1$, the larger number of $\left\langle N_{\rm
  sat}\right\rangle$ is allowed in the model (see the factor of $1/2$
in the Zheng model).  
Furthermore, since the Zheng model has 
$\left\langle N_{\rm cen}\right\rangle =1$ at the high mass end, 
it would make sense for the Zheng model to have the lower number of 
$\left\langle N_{\rm sat}\right\rangle$ when the total number density is fixed.
Nevertheless, the two $N(M)$'s are consistent with each other within $2-\sigma$.
The best-fitting correlation function is shown as the red dashed curve
in the upper and lower panels of figure \ref{fig:result_xi_contreras}
for $z=1.19$ and $z=1.47$, respectively.  
As expected, the overall shape of
the correlation function is very similar with that of the Geach model
depicted by the red solid curve.
However, one can see that the Geach model shows a better fit to the 
measured $w(\theta)$ at the smallest scales, which would imply that 
the ELGs in our sample prefer the HOD model which allows 
massive halos to have no central ELG. 

The posterior distribution of the derived physical parameters is shown
by the function colored orange in figure
\ref{fig:result_params_b_m_n_f}.  Since the number density is
primarily constrained by our prior, the two models predict the same
values.  We find that the effective bias $b_{\rm eff}$ and halo mass
$\log{M_{\rm eff}}$ determined from the Zheng HOD model are also
consistent with those from the Geach model within the statistical
uncertainties.  On the other hand, the satellite fraction is derived
to be $f_{\rm sat} =0.102^{+0.013}_{-0.012}$ and
$0.093^{+0.012}_{-0.011}$ at $z=1.19$ and $1.47$, respectively.  These
values are slightly lower than those determined based on the Geach HOD
model, respectively $0.158^{+0.114}_{-0.047}$ and
$0.159^{+0.109}_{-0.049}$, though the differences are not significant.
This again reflects the fact that the Geach model adopts a more
flexible parameterization and the larger amplitude of $\left\langle
N_{\rm sat}\right\rangle$ is allowed.  Hence the 1-d posterior of
$f_{\rm sat}$ obtained from the Geach model has a long tail toward the
high $f_{\rm sat}$ end.

Overall, the differences of halo parameters for [OII] emitters
determined from the Geach and Zheng HOD models are not significant
compared to the current statistical uncertainties.  However, such
small differences need to be carefully treated in the future
clustering analysis ELGs from large galaxy surveys such as the Subaru
PFS, DESI, and Euclid surveys.

\section{Summary}\label{sec:conclusion}

In this paper we have studied the clustering of emission line galaxies
at $z>1$ and physical properties of dark matter halos which host them.
For this purpose we used [OII] emitters detected at $z=1.19$ and
$1.47$ using two narrow band filters, {\it NB816} and {\it NB921},
respectively, in the Subaru HSC survey \citep{Hayashi:2020}.  We then
measured the angular correlation functions of 8302 ($z=1.19$) and 9578
($z=1.47$) of [OII] emitters.  Using a simple model of the
nonlinear correlation function with the linear galaxy bias factor, we
measured the bias of [OII] emitters as $b=1.61^{+0.13}_{-0.11}$ and
$2.09^{+0.17}_{-0.15}$ at $z=1.19$ and $1.47$, respectively. 
We also found that the bias monotonically increases with the line luminosity.

We, for the first time, performed an HOD analysis for the measured
correlation functions of [OII] emitters based on a model developed to
describe the population of galaxies selected by a star forming rate
\citep{Geach:2012}.  We varied eight parameters simultaneously with
only a few priors.  Nevertheless, the two HOD parameters related to
the host halo mass have been well constrained given 
the large sample from the HSC survey.

Based on the constrained HOD parameters, we have derived parameters
which describe properties of the host halos of [OII] emitters, such as
the effective bias, effective halo mass and satellite fraction.  The
effective biases are determined as $b_{\rm
  eff}=1.701^{+0.083}_{-0.110}$ and $1.981^{+0.072}_{-0.068}$ at
$z=1.19$ and $1.47$, respectively.  They are consistent with the
linear bias values determined based on a simple nonlinear dark matter
model.  The effective halo masses and satellite fractions are derived
as $\log{M_{\rm eff}/(h^{-1}M_\odot)} \simeq 12.7~ (12.6)$ and $f_{\rm
  sat}\simeq 0.16~(0.16)$ at $z=1.19~(1.47)$.  They are consistent
with the result of \citet{Guo:2019} who studied the relation among
$f_{\rm sat}$, $M_{\rm eff}$ and the stellar mass using the ELG sample
from the eBOSS survey.
Furthermore, the determined effective halo masses are in good
agreement with previous studies of similar tracers with the
mass-assembly history with $M(z=0)=1.5\times 10^{13}h^{-1}M_\odot$.

As stated in section \ref{sec:introduction}, ELGs will be a main
tracer in large galaxy surveys at high redshifts.  Particularly, [OII]
emitters will be targeted by many ongoing/future surveys such as the
Subaru PFS survey \citep{Takada:2014}.  The result presented in this
paper is useful to construct a mock catalog for cosmological analysis.
The physical properties of halos hosting [OII] emitters revealed in
this paper can also be applicable to the Subaru PFS, DESI, and other
forthcoming surveys.

\begin{figure*}
\begin{center}
\vspace{-.8cm}
\FigureFile(172mm,172mm){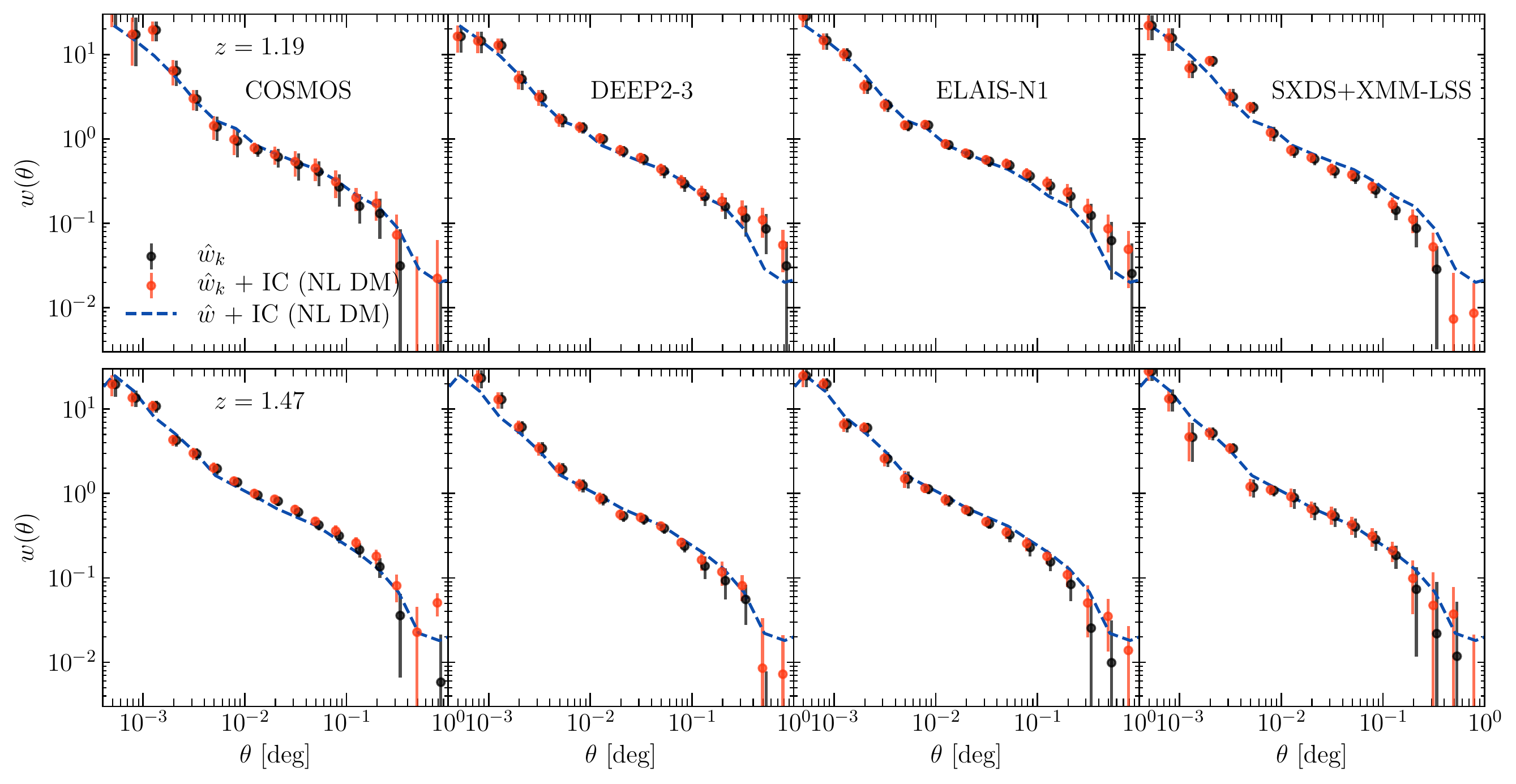}
\end{center}
\caption{Angular correlation functions of [OII] emitters of each of
  the four fields measured at $z=1.19$ (top) and $z=1.47$ (bottom).
  In each column, the measurement from each of the individual four
  fields is presented.  The black points are the raw measurement
  without the correction of the integral constraints, $\hat{w}_k$
  ($1\leq k \leq 4$).  The red points are after the integral
  constraint (denoted by IC in the legend) is taken into account with
  the linearly-biased dark matter model, $\hat{w}_k+(1-f_{\rm fake})^2w_{\Omega,k}$,
  with the best-fitting parameters obtained
  for the combined field, $(b,f_{\rm fake})=(1.60, 0.140)$ $(z=1.19)$
  and $(b,f_{\rm fake})=(2.08, 0.140)$ $(z=1.47)$.  For the reference,
  the angular correlation function measured for the combined field,
  $\hat{w}+(1-f_{\rm fake})^2w_\Omega$, is shown as the blue dashed
  curve.  }
\label{fig:w_individual}
\end{figure*}

In this analysis, both the shape of the angular correlation function
and number density are constrained by the HOD modeling.  However,
there is a claim that in 3-dimensional analysis the HOD would fail to
explain the redshift-space clustering and number density
simultaneously \citep{Reid:2014,Saito:2016}.  To investigate this, we
need a spectroscopic survey, and we can do it with the PFS, DESI or
{\it Euclid} surveys.

The NB data of the HSC-SSP PDR2 provide emission lines of not only
[OII] but also H$\alpha$ and [OIII], ranging from $z\sim 0.4$ to
$z\sim 1.6$ \citep{Hayashi:2020}.  Redshift evolution of physical
properties such as the halo mass for ELGs using these multiple
emission lines will be studied in our future work.


\begin{table*}[bt!]
\caption{Priors and constraints of the HOD parameters for Geach and Zheng models obtained using the clustering information only${}^*$
}
\begin{center}
\begin{tabular}{ll | l | l}
\multicolumn{4}{c}{Geach HOD model} \\
\hline 
                   &          & \multicolumn{1}{c |}{{\it NB816}} & \multicolumn{1}{c}{{\it NB921}} \\
Parameter  & Prior & Posterior PDF & Posterior PDF \\
\hline 
$\log{M_{\rm c}} / (h^{-1}M_\odot)$     &  None               & $11.50^{+0.72}_{-0.28}$     & $11.70^{+ 0.25} _{- 0.20}$ \\  
$\log{M_{\rm min}} / (h^{-1}M_\odot)$ & None                & $11.40^{+1.25}_{-1.00}$    & $11.89^{+0.76}_{-1.36}$ \\ 
$\sigma_{\log{M}}$                              & $[0,1]$             & $0.27^{+0.33}_{-0.17}$      & $0.22^{+ 0.16}_{- 0.13}$  \\
$\alpha$                                               & $1.00\pm 0.20$  & $0.96^{+0.11}_{-0.09}$       & $1.10^{+ 0.13}_{- 0.13}$ \\
$F_c^A$                                               & $[0, 0.5]$         & $0.22^{+0.14}_{-0.07}$        & $0.17^{+ 0.18}_{- 0.06}$  \\
$F_c^B$                                               & $[0,1]$          & $0.56^{+0.32}_{-0.39}$        & $0.61^{+ 0.27}_{- 0.34}$  \\
$F_s$                                                  & $[0, 1]$            & $0.54^{+0.32}_{-0.30}$      & $0.52^{+ 0.32}_{- 0.34}$  \\
$f_{\rm fake}$                                      & $0.140\pm 0.060$  & $0.124^{+0.054}_{-0.051}$ & $0.091^{+ 0.054}_{- 0.046}$  \\
\hline 
Inferred quantity & Measurement & Posterior PDF  &  Posterior PDF \\
\hline
$\log{n_g / (h^{-1}{\rm Mpc})^{-3}}$ & \multicolumn{1}{c |}{$\cdots$} &  $-1.47^{+0.89}_{-0.80}$ & $-1.96^{+1.18}_{-0.42}$  \\
$f_{\rm sat}$ & \multicolumn{1}{c |}{$\cdots$}                                        & $0.58^{+0.33}_{-0.43}$  & $0.38^{ + 0.54}_{ - 0.23}$ \\
$b_{\rm eff}$ & \multicolumn{1}{c |}{$\cdots$}                                       & $1.64^{+0.12}_{-0.10}$  & $1.94^{+0.13}_{-0.11}$ \\
$\log{M_{\rm eff}} / (h^{-1}M_\odot) $ & \multicolumn{1}{c |}{$\cdots$}  & $12.70^{+0.10}_{-0.09}$  & $12.63^{+0.12}_{-0.10}$  \\
\hline \\
\multicolumn{4}{c}{Zheng HOD model} \\
\hline 
& & \multicolumn{1}{c|}{{\it NB816}}  &  \multicolumn{1}{c}{{\it NB921}}   \\ 
Parameter  & \multicolumn{1}{l |}{Prior} & Posterior PDF & Posterior PDF \\
\hline 
$\log{M_{\rm min}} / (h^{-1}M_\odot)$ & \multicolumn{1}{l |}{None}         & $12.01^{+0.62}_{-0.61}$ &         $11.85^{+ 0.56}_{-0.26}$ \\ 
$\log{M_1} / (h^{-1}M_\odot)$             & \multicolumn{1}{l |}{None}       & $12.97^{+0.43}_{-0.64}$  &    $12.91^{+0.36}_{-0.31}$\\ 
$\sigma_{\log{M}}$                              & \multicolumn{1}{l |}{$[0,1]$}            & $0.58^{+0.30}_{-0.41}$ &            $0.32^{+0.41}_{-0.22}$  \\
$\alpha$                                               & \multicolumn{1}{l |}{$1.00\pm 0.20$}     & $0.94^{+0.11}_{-0.11}$ &         $1.02^{+0.11}_{-0.11}$ \\
$f_{\rm fake}$                                      & \multicolumn{1}{l |}{$0.140\pm 0.060$} & $0.123^{+0.060}_{-0.063}$  &  $0.077^{+ 0.055}_{- 0.047}$  \\
\hline 
Inferred quantity & \multicolumn{1}{l |}{Measurement}  & Posterior PDF & Posterior PDF \\
\hline
$\log{n_g / (h^{-1}{\rm Mpc})^{-3}}$ & \multicolumn{1}{c |}{$\cdots$}    & $-2.24^{+0.56}_{- 0.44}$ &           $-2.27^{+0.28}_{-0.38}$ \\
$f_{\rm sat}$ & \multicolumn{1}{c |}{$\cdots$}                                       & $0.113^{+0.100}_{-0.044}$ &     $0.106^{+ 0.047}_{-0.040}$ \\
$b_{\rm eff}$ & \multicolumn{1}{c |}{$\cdots$}                                       & $1.64^{+ 0.15}_{-0.15}$ &           $1.91^{+0.13}_{-0.11}$ \\
$\log{M_{\rm eff}} / (h^{-1}M_\odot) $ & \multicolumn{1}{c |}{$\cdots$} & $12.633^{+0.102}_{-0.107}$ & $12.534^{+0.091}_{-0.077}$  \\
\hline
\end{tabular}
\end{center}
\label{tab:hod_nong}
\begin{tabnote}
$*$ In the ``Prior'' column the ranges specified in brackets are for
  uniform priors while the others we quote the mean and standard
  deviation of the Gaussian priors. In the
  column of ``Posterior PDF'', the central value is a median and the
  error means $16-84$ percentiles after other parameters are
  marginalized over.
\end{tabnote}
\end{table*}

\section*{Acknowledgments }
TO thanks the members of the Subaru PFS Cosmology Working Group,
particularly Masahiro Takada, Eiichiro Komatsu, and Ryu Makiya for useful
correspondences during the regular telecon. 
TO is grateful to Shogo Ishikawa for discussion.
We also thank the anonymous referee for the careful reading and suggestions.
TO acknowledges support
from the Ministry of Science and Technology of Taiwan under Grants
No. MOST 109-2112-M-001-027- and the Career Development Award,
Academia Sinica (AS-CDA-108-M02) for the period of 2019-2023.  KO is
supported by JSPS Overseas Research Fellowships.  This work is based
on data collected at Subaru Telescope, which is operated by the
National Astronomical Observatory of Japan.

The Hyper Suprime-Cam (HSC) collaboration includes the astronomical
communities of Japan and Taiwan, and Princeton University. The HSC
instrumentation and software were developed by the National
Astronomical Observatory of Japan (NAOJ), the Kavli Institute for the
Physics and Mathematics of the Universe (Kavli IPMU), the University
of Tokyo, the High Energy Accelerator Research Organization (KEK), the
Academia Sinica Institute for Astronomy and Astrophysics in Taiwan
(ASIAA), and Princeton University. Funding was contributed by the
FIRST program from the Japanese Cabinet Office, the Ministry of
Education, Culture, Sports, Science and Technology (MEXT), the Japan
Society for the Promotion of Science (JSPS), Japan Science and
Technology Agency (JST), the Toray Science Foundation, NAOJ, Kavli
IPMU, KEK, ASIAA, and Princeton University.

\appendix
\section{Correlation functions in individual fields}\label{sec:w_individual}

In order to confirm the consistency of the angular correlation
functions among different fields, we present the measurement from each
of the individual four fields, $\hat{w}_k$, as the black points in
Figure \ref{fig:w_individual}.
The square roots of the diagonal components of the covariance matrix,
$C_{k,ii}^{1/2}$, are shown as the error bars.
The results for $z=1.19$ and $z=1.47$ are shown in the upper and lower
panels, respectively.  In order to compare the correlation functions
between different fields, we need to take into account the correction
of the integral constraint because the survey volume of each field is
different.
To estimate the integral constraint for $k$-th field, $w_{\Omega,k}$,
the random-random count for the field, $RR_k$, is used as $RR$ in
equation (\ref{eq:ic_rr}), and
\be
w_{\Omega,k}(\Theta) = \frac{\sum_i{w(\theta_i;\Theta)RR_k(\theta_i)}} {\sum_i{RR_k(\theta_i)}}. \label{eq:ic_rr_k}
\ee
For $w(\theta_i;\Theta)$ in equation (\ref{eq:ic_rr_k}), we use the
best-fitting linearly-biased dark matter model for the combined field
obtained in section \ref{sec:nl_matter}, $(b,f_{\rm
  fake})=(1.60,0.140)$ for $z=1.19$ and $(b,f_{\rm
  fake})=(2.08,0.140)$ for $z=1.47$.
The red points show the result with the integral constraint
correction, $\hat{w}_k+(1-f_{\rm fake})^2 w_\Omega$.  For comparison,
the measurement from all four fields with the integral
constraint calculated using the same model is plotted as the blue
dashed curve. One can see that the measurement from each field is
largely consistent with each other.  Thus, throughout this paper we
perform the statistical analysis only for the data combined over all
the four fields.

\section{Constraints on HOD parameters using clustering only}\label{sec:hod_wo_nobs}
Given an HOD model for a certain galaxy population, 
the average number density is calculated by equation (\ref{eq:ng_hod}). 
One therefore needs to simultaneously analyze the observed correlation function and 
number density to constrain the HOD model, as performed in section \ref{sec:constraint}. 
In this appendix, we present the constraints on the HOD parameters based on Geach and Zheng models 
without using the information of the measured number density of [OII] emitters for readers who are interested in the constraining power 
coming from the clustering only (see, e.g., \cite{Matsuoka:2011,Martinez-Manso:2015} for a similar attempt). 

Table \ref{tab:hod_nong} shows these results without using the abundance information. 
The halo mass parameters are strongly constrained by the abundance of [OII] emitters. 
The constraints on these parameters are thus not as tight as those presented in tables \ref{tab:hod_geach} and \ref{tab:hod_zheng}, as expected:
the errors on $M_{\rm c}$ and $M_{\rm min }$ constraints without the abundance information in the Geach model are $\sim 1.5$ and $\sim 3$ times 
larger than those with the simultaneous analysis of clustering and abundance, and the errors on $M_{\rm min}$ and $M_1$ constraints without the abundance information in
the Zheng model are $\sim 3$ and $\sim 4$ times larger. 
There are not significant changes in the constraints of the other HOD parameters. 
Interestingly, even though we do not use the number density to constrain the HOD parameters, 
the posterior of the number density is consistent with the observed one. 


\end{document}